\documentclass[manuscript]{acmart}

\AtBeginDocument{%
  }

\setcopyright{acmcopyright}
\copyrightyear{2023}
\acmYear{2023}
\acmDOI{XXXXXXX.XXXXXXX}

\acmConference[RecSys '23]{Make sure to enter the correct
  conference title from your rights confirmation emai}{September 18--22,
  2023}{Singapore}
\acmPrice{15.00}
\acmISBN{978-1-4503-XXXX-X/18/06}

\usepackage{multirow}
\newcommand{\framework}{ColdGPT}
\usepackage{array}
\newcolumntype{P}[1]{>{\centering\arraybackslash}p{#1}}
\usepackage{graphicx}
\usepackage[caption=false]{subfig}
\usepackage{natbib}
\begin{document}

\title{Multi-task Item-attribute Graph Pre-training for Strict Cold-start Item Recommendation}


\author{Yuwei Cao}
\affiliation{%
  \institution{University of Illinois Chicago}
  \city{Chicago}
  \country{USA}}
\email{ycao43@uic.edu}

\author{Liangwei Yang}
\affiliation{%
  \institution{University of Illinois Chicago}
  \city{Chicago}
  \country{USA}}
\email{lyang84@uic.edu}

\author{Chen Wang}
\affiliation{%
  \institution{University of Illinois Chicago}
  \city{Chicago}
  \country{USA}}
\email{cwang266@uic.edu}

\author{Zhiwei Liu}
\affiliation{%
  \institution{Salesforce AI}
  \country{USA}}
\email{zhiweiliu@salesforce.com}

\author{Hao Peng}
\authornote{This is the corresponding author.}
\affiliation{%
  \institution{Beihang University}
  \city{Beijing}
  \country{China}}
\email{penghao@buaa.edu.cn}

\author{Chenyu You}
\affiliation{%
  \institution{Yale University}
  \city{New Haven}
  \country{USA}}
\email{chenyu.you@yale.edu}

\author{Philip S. Yu}
\affiliation{%
  \institution{University of Illinois Chicago}
  \city{Chicago}
  \country{USA}}
\email{psyu@uic.edu}
\renewcommand{\shortauthors}{Cao et al.}

\begin{abstract}
Recommendation systems suffer in the \textit{strict cold-start (SCS)} scenario, where the user-item interactions are entirely unavailable. The well-established, dominating identity (ID)-based approaches completely fail to work. Cold-start recommenders, on the other hand, leverage item contents (brand, title, descriptions, etc.) to map the new items to the existing ones. However, the existing SCS recommenders explore item contents in \textit{coarse-grained} manners that introduce noise or information loss.
Moreover, informative data sources other than item contents, such as users' purchase sequences and review texts, are largely ignored. 
In this work, we explore the role of the \textit{fine-grained} item attributes in bridging the gaps between the existing and the SCS items and pre-train a knowledgeable item-attribute graph for SCS item recommendation. 
Our proposed framework,~\framework, models item-attribute correlations into an item-attribute graph by extracting fine-grained attributes from item contents. \framework~then transfers knowledge into the item-attribute graph from various available data sources, i.e., item contents, historical purchase sequences, and review texts of the existing items, via multi-task learning. 
To facilitate the positive transfer,~\framework~designs specific submodules according to the natural forms of the data sources and proposes to coordinate the multiple pre-training tasks via unified alignment-and-uniformity losses.
Our pre-trained item-attribute graph acts as an implicit, extendable item embedding matrix, which enables the SCS item embeddings to be easily acquired by inserting these items into the item-attribute graph and propagating their attributes' embeddings. We carefully process three public datasets, i.e., Yelp, Amazon-home, and Amazon-sports, to guarantee the SCS setting for evaluation. Extensive experiments show that~\framework~consistently outperforms the existing SCS recommenders by large margins and even surpasses models that are pre-trained on 75 - 224 times more, cross-domain data on two out of four datasets. Our code and pre-processed datasets for SCS evaluations are publicly available to help future SCS studies.
\end{abstract}



\received{21 April 2023}

\maketitle

\section{Introduction}

\begin{figure}
    \centering
    \subfloat[\centering Amazon-home purchases in 2017]{{\includegraphics[width=7.2cm]{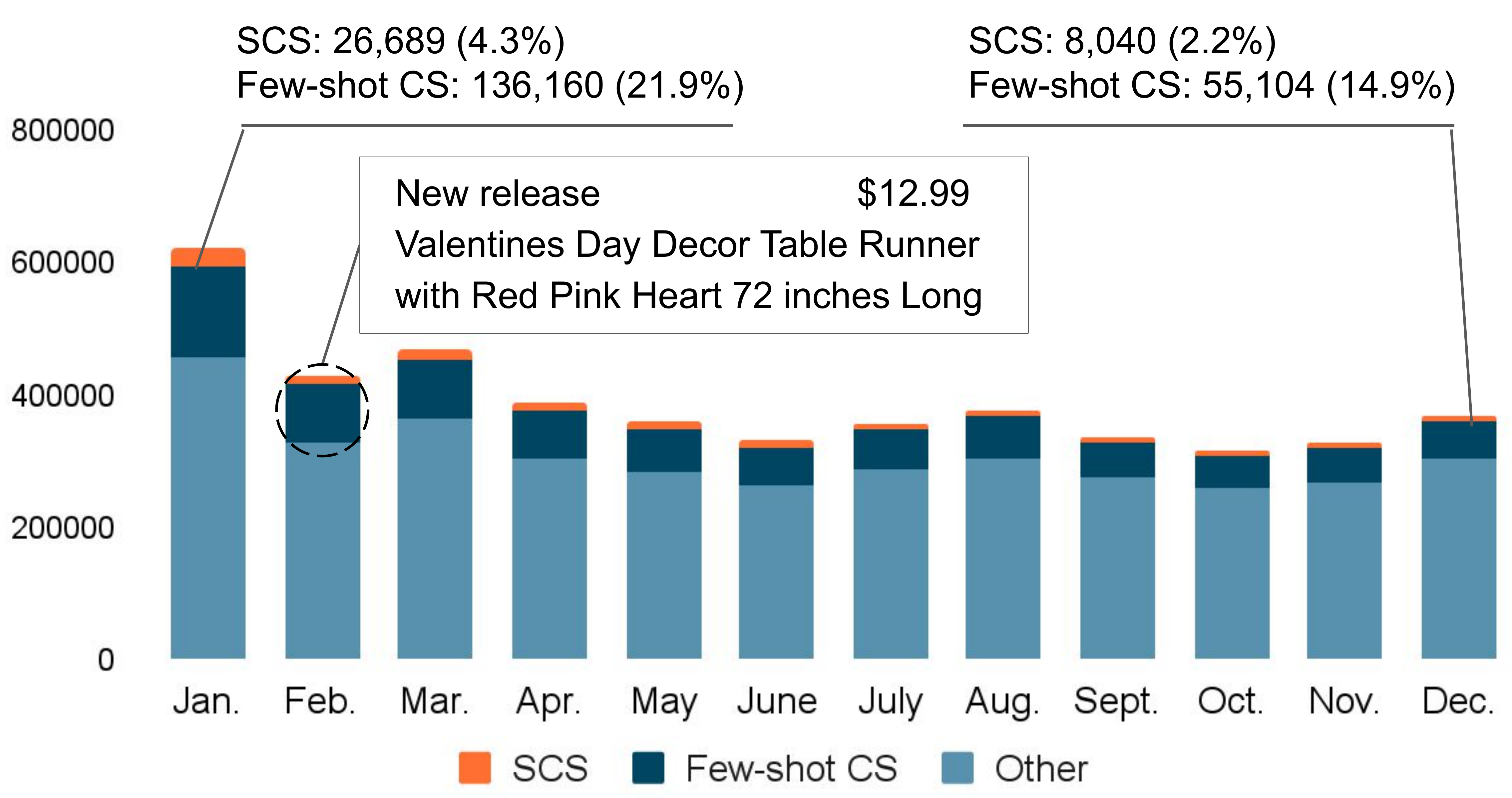} }}
    \subfloat[\centering An example of our proposed item-attribute graph]{{\includegraphics[width=7.7cm]{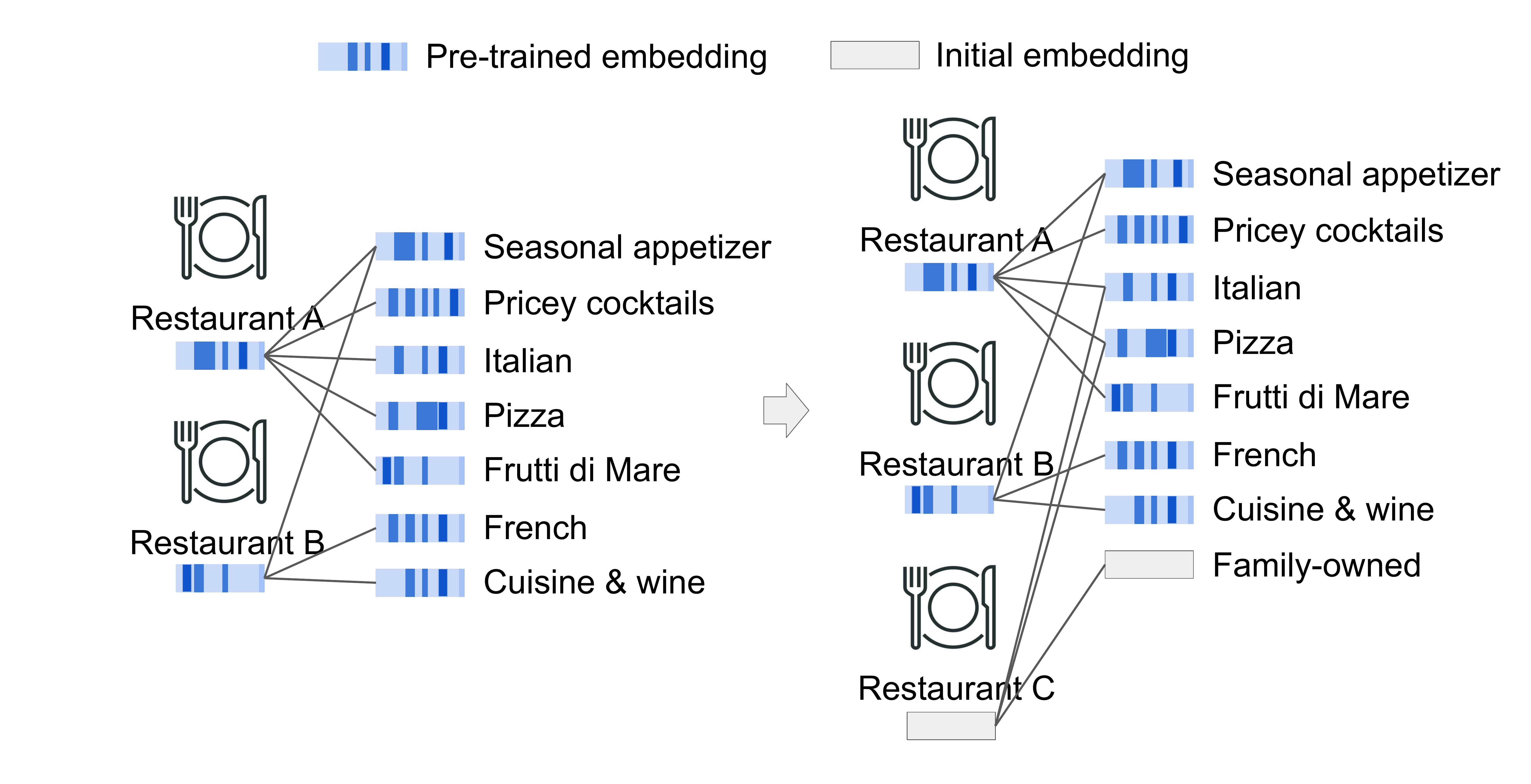} }}
    \caption{The distribution of Amazon-home purchases in 2017 (a) and an example of our proposed item-attribute graph (b). In (a), `SCS', `Few-shot CS', and `Other' count the purchases of the SCS, few-shot cold-start (items that have been bought > 1 and $\leq$ 20 times), and the other items, respectively. In (b), the LHS shows a pre-trained item-attribute graph and the RHS shows the updated item-attribute graph with SCS restaurant C and its new attribute, `Family-owned', inserted.}
    \label{fig:cold_start_item_attribute}
\vspace{-1em}
\end{figure}


Recommendation systems suffer in cold-start scenarios where the well-established, dominating identity (ID)-based approaches deteriorate \cite{wheretogo2023, wu2021towards}. The problem is especially severe in the zero-shot, \textit{strict cold-start (SCS)} \cite{zheng2021cold} setting. In this work, we propose to address SCS item recommendation, i.e., recommend new items that are unseen during training without relying on any historical rating records of these new items. This is to be differentiated from the \textit{few-shot cold-start} item recommendation \cite{wu2021towards}, in which the new items' rating records, though inaccessible during training and/or pre-training, are available for inference and/or fine-tuning.

Recommending cold-start items is of great importance. As shown in Figure \ref{fig:cold_start_item_attribute} (a), SCS and few-shot cold-start items (e.g., festival new releases) account for up to $4\%$ and $22\%$ of the monthly Amazon-home purchases in the year 2017, respectively.
Cold-start recommendation is an essential task, yet is under-explored \cite{wu2021towards}. 
Few studies address SCS item recommendation, as it is extremely difficult: without rating records, we have little knowledge about new items, merely \textit{item contents}, i.e., auxiliary features such as price, brand, descriptions, etc. Intrinsically, one would resort to transferring knowledge from the warm-start items, which are associated with abundant information (historical rating scores and review texts, in addition to item contents). 
As such, existing SCS recommenders \cite{volkovs2017dropoutnet, barkan2019cb2cf, zhu2020recommendation, wei2021contrastive, zhao2022improving, li2019zero, qian2020attribute} strive to reconstruct the preference embeddings of the users and items from their content embeddings.
In addition, owing to the rapid advance of NLP, recent methods such as UniSRec \cite{hou2022towards} and P5 \cite{geng2022recommendation} adapt pre-trained language models (PLMs) \cite{devlin2018bert, raffel2020exploring} for recommendation and are capable of accommodating SCS items. 
However, existing methods merely explore the available data sources in \textit{coarse-grained} manners. 
Specifically, methods that reconstruct the preference embeddings ignore both the review texts and the item correlations contained in the users' rating sequences. UniSRec \cite{hou2022towards} simply concatenates item contents and truncates the information due to the input length limitations of PLMs. 
P5 \cite{geng2022recommendation} distorts various data sources, including the numerical rating sequences, into natural language sequences to fit its PLM backbone \cite{raffel2020exploring}. 
Such coarse-grained processing tends to mix useful information with noise and hinders the preference characterization ability of models.

Hence, this paper investigates how to leverage the information for SCS items in a \textit{fine-grained} manner.
Specifically, we study how to effectively harness the item \textit{attributes}, i.e., phrases that describe the items, to characterize proper item information for the recommendation, especially for those SCS items. 
We observe that though new items are ever-emerging, their attributes are relatively stable. We demonstrate in Table~\ref{dataset} that $2,867$ test items in the Home-SCS dataset introduce only one $1$ new attribute. Therefore, if we train a recommendation model via the attributes of items, it is also able to recommend SCS items.
In Figure~\ref{fig:cold_start_item_attribute} (b), for example, a user that likes restaurant A may be interested in a new restaurant C that shares the same style (`Italian') and serves similar foods (`Pizza'). 
Based on this intuition, we propose to pre-train our model on a bipartite item-attribute graph for SCS item recommendation. 
The goal is to transfer item-related knowledge from available data resources into attribute embeddings. 
Informative SCS item embeddings can then be easily acquired upon the arrival of these items — insert the SCS items along with the new attributes into the item-attribute graph and propagate their neighbors' embeddings. 
For example, in Figure \ref{fig:cold_start_item_attribute} (b), restaurant C's embedding is obtained by fusing the embeddings of its attributes, i.e., `Italian', `Pizza', and `Family-owned'. The first two contain pre-trained knowledge related to the existing restaurants while the last one introduces information unique to restaurant C. 
Arguably, our proposed item-attribute graph is an implicit, extendable counterpart of the traditional explicit, fixed item ID embedding matrix in the SCS setting: it allows encoding items with only their attributes.

Pre-training a knowledgeable item-attribute graph requires properly addressing three challenges. \textbf{1) Construction of the item-attribute graph}. 
On the one hand, mining fine-grained item attributes from the item contents is non-trivial. The item contents can be long and noisy, containing all-caps sentences, URLs, special characters, etc. We carefully pre-process them and extract noun phrases (which better describe the items than the raw item contents, as demonstrated in Section \ref{sec:construction}) to serve as high-quality item attributes. On the other hand, review texts reflect users' subjective perception of the items' attributes and serve as valuable supplements to the objective item contents. However, due to the high volume and complexity of the review texts, using them in straightforward manners is costly and ineffective (empirically verified in Section \ref{sec:experiment}). Inspired by studies in phrase-level sentiment analysis \cite{zhang2014users}, we extract phrases with sentiments and model the item correlations reflected by them. For example, in Figure \ref{fig:cold_start_item_attribute} (b), subjective descriptions such as `Seasonal appetizer' helps to reveal the subtle correlations between restaurants A and B. 

\textbf{2) Transferring knowledge from various available data sources}, i.e., item contents, historical rating scores, and review texts. 
To preserve the above information, we design submodules according to their natural forms for further exploration. Specifically, the item contents, after modeled into the aforementioned item-attribute graph, are then embedded with PLM and graph neural network (GNN) \cite{kipf2016semi, velivckovic2017graph} to capture the underlying natural language semantics and higher-order structural information. The historical rating scores are modeled with a Transformer \cite{vaswani2017attention}-based sequential submodule to extract item-item correlations. The review texts are mined in a manner similar to the item contents but with a separate submodule.

\textbf{3) Simultaneously incorporating all knowledge.} We conduct multi-task learning (MTL) \cite{caruana1997multitask} with self-supervised pre-training tasks specifically designed for these sub-modules to transfer knowledge into the item-attribute graph. 
We observe negative transfer \cite{caruana1997multitask} between the tasks, which leads to sub-optimal performance when adopting the commonly used loss terms for the different sub-modules (detailed in Section \ref{sec:ablation_study}). 
We hypothesize that negative knowledge transfer results from the inconsistency of different objective functions. 
Thus, we propose to integrate knowledge from different tasks in a unified manner. 
Specifically, we propose to coordinate embeddings learned from different tasks via consistent \textit{alignment} \cite{wang2020understanding} losses.
In this way, knowledge from multiple tasks is effectively shared. 
However, enforcing the alignment of multi-task knowledge without regularization induces the embedding collapse problem \cite{wang2020understanding,jing2021understanding} as the optimization process degenerates to a trivial solution. 
Hence, we devise a multi-task \textit{uniformity} \cite{wang2020understanding} regularization objective, which enables those embeddings to be uniformly distributed on a common hyper-sphere.
As such, knowledge can be effectively fused and positively transferred. 
Differing from recent studies \cite{wang2022towards} that leverage alignment-and-uniformity for collaborative filtering, we are the first work to demonstrate the extraordinary ability of multi-task alignment-and-uniformity in knowledge transferring. 
Additionally, we address the discrepancies among various pre-training tasks and investigate how to simultaneously characterize the item-attribute, item-item, and item-review correlations via a multi-task training paradigm.

Our proposed unified task losses greatly benefit the multi-task pre-training, as demonstrated in Section \ref{sec:ablation_study}. To further facilitate the positive transfer, we carefully tune hyperparameters including the task weights, learning rate, and regularization weight, as recent MTL studies \cite{kurin2022defense, xin2022current} suggest that such an approach essentially outperforms the ad-hoc MTL methods \cite{javaloy2021rotograd, yu2020gradient}.
We name our proposed framework as an item-attribute graph pre-training based SCS item recommender, i.e., \framework~(GPT stands for graph pre-training, which is to be differentiated from generative pre-trained transformer \cite{brown2020language}). We carefully process three public datasets, i.e., Yelp, Amazon-home, and Amazon-sports, to guarantee the SCS setting for~\framework's evaluation. Extensive experiments show that~\framework~outperforms the existing cold-start recommenders consistently by large margins and even surpasses models that are pre-trained on high-volume, cross-domain data on two out of four datasets.

The contributions of our work are:
1) we address item recommendation in the SCS setting, which is extremely challenging and under-explored. Compared to the previous SCS methods, we consider more information, i.e., sequential item correlations and review texts, in addition to item contents.
2) we explore item contents in a fine-grained manner like no previous recommenders and explicitly model the item attributes to bridge the gaps between the existing and the SCS items. Our pre-trained item-attribute graph serves as an implicit, extendable counterpart of the traditional item ID embedding matrix in the SCS setting and allows us to easily encode items using their attributes.
3) we effectively transfer knowledge from various data sources, i.e., item contents, rating scores, and review texts of the existing items, into the item-attribute graph. Our proposed framework,~\framework, adopts submodules that are specifically designed with respect to the natural forms of various data sources. 
The multi-task pre-training paradigm of~\framework~is the first to demonstrate the extraordinary ability of multi-task alignment-and-uniformity in positive knowledge transferring.
4) we evaluate \framework~in the SCS setting, which is either overlooked or under-explored by the previous work in their evaluations. Extensive experiments show the superior performance of~\framework~over various baselines and the effectiveness of~\framework's components. To help future SCS studies, our code and datasets carefully processed to keep the SCS setting are publicly available\footnote{\url{https://github.com/YuweiCao-UIC/ColdGPT}}.

\section{Methodology}
\begin{figure*}[t]
\centering
\includegraphics[width = 15cm]{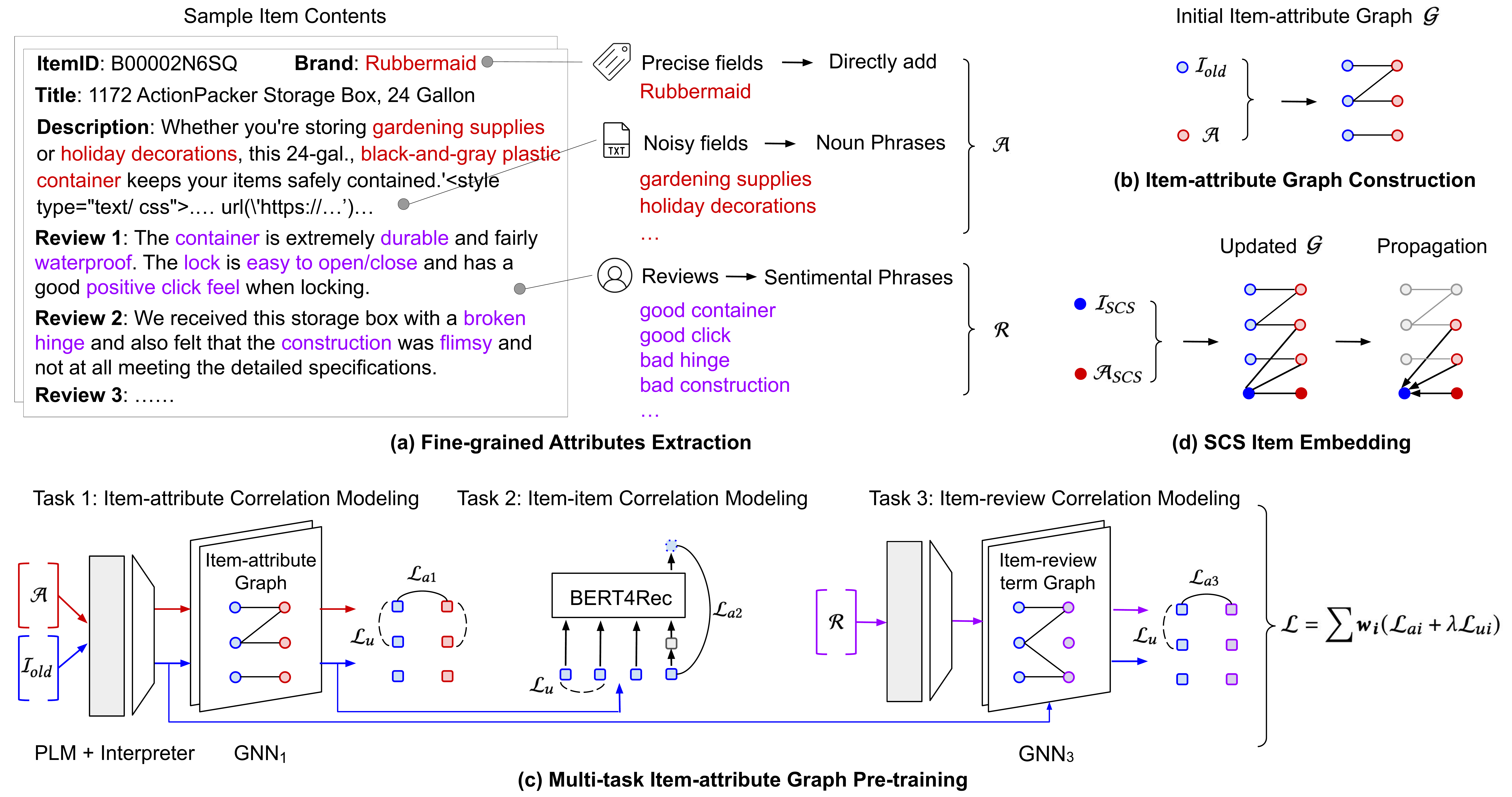}
\caption{\label{fig:framework} The overall architecture of our proposed~\framework~framework. The blue, red, and purple nodes indicate the items, attributes, and review terms, respectively. (a) shows the extraction of the fine-grained attributes. (b) shows the construction of the item-attribute graph. (c) shows the multi-task pre-training of the item-attribute graph, in which the parameters of the PLM (grey rectangle) are fixed. (d) shows SCS item embedding with the pre-trained item-attribute graph.}
\vspace{-1em}
\end{figure*}

In this section, we pre-train an item-attribute graph that enables efficient and effective inference of SCS item embedding for solving SCS item recommendation.
Figure \ref{fig:framework} shows the overall architecture of the proposed~\framework~framework. 
We first formulate the task in Section \ref{sec:problem_formalization}. Section \ref{sec:construction} presents the construction of the item-attribute graph. Section \ref{sec:pre-training} illustrates how the item-attribute graph is pre-trained by transferring knowledge from various available data sources. Section \ref{sec:inference} introduces how SCS item embeddings can be easily acquired using the pre-trained item-attribute graph.
\subsection{SCS Item Recommendation}
\label{sec:problem_formalization}
Let $\mathcal{U}$, $\mathcal{I}_{old}$, and $\mathcal{I}_{SCS}$ denote the user, the existing item, and the SCS item set, respectively, and $\mathcal{I}_{old}\cap\mathcal{I}_{SCS}=\emptyset$. In the task of SCS item recommendation, we are given 1) $\mathcal{Y}_{old} = \{(u,i)|u\in\mathcal{U}, i\in\mathcal{I}_{old}\}$, which is a set of observed user-item interactions (optionally, each interaction comes with a textual review) and 2) $\mathcal{C} = \{\mathcal{C}_i|i\in\mathcal{I}_{old}\cup\mathcal{I}_{SCS}\}$, which is a set of item contents and the item contents $\mathcal{C}_i$ of item $i$ include \textit{title}, \textit{brand}, \textit{category}, \textit{description field}, etc., depending on the specific dataset. Our goal is to learn a model $f_\theta$ that recommends $\mathcal{I}_{SCS}$ to $\mathcal{U}$, i.e., $f_\theta(\mathcal{C},\mathcal{Y}_{old}) = \hat{\mathcal{Y}}_{SCS} = \{(u,i)|u\in\mathcal{U}, i\in\mathcal{I}_{SCS}\}$ and we want the recommendations to be as accurate as possible. I.e., $maximize(|\hat{\mathcal{Y}}_{SCS} \cap \mathcal{Y}_{SCS}|)$, where $\mathcal{Y}_{SCS} = \{(u,i)|u\in\mathcal{U}, i\in\mathcal{I}_{SCS}, u \text{ subsequently interacts with } i\}$.
\subsection{Item-attribute Graph Construction}
\label{sec:construction}
The main challenge in constructing the proposed item-attribute graph is determining the attributes given the item contents, i.e., fields such as \textit{brand}, \textit{category}, \textit{description}, etc. Considering their essential role as media in bridging the items, the attributes should only contain terms that reflect the property of the items.
As shown in Figure \ref{fig:framework} (a), fields such as brand are short and precise. Therefore, they can be directly added to the set of attributes, denoted as $\mathcal{A}$. For long, noisy fields, however, pre-processing is necessary.
We observe that compared to the raw sentences, the noun phrases better describe the items with less irrelevant information. For example, the raw description field of the sample item contents in Figure \ref{fig:framework} (a) contains words that are irrelevant to the item, such as `whether' and `you're'. There is also noise like hashtags and URLs. In contrast, the noun phrases contained in the description field, i.e. `gardening supplies', `holiday decorations', and `black-and-gray plastic container', precisely describe the item with much less redundant information.
Therefore, we extract noun phrases to serve as item attributes and add these attributes into $\mathcal{A}$. 
We first filter out the noise and non-human-readable contents such as URLs, HTML tags, and special characters, and then leverage the NLP toolkit \footnote{\url{https://textblob.readthedocs.io/en/dev/}} for noun phrases extraction. As shown in Figure \ref{fig:framework} (b), the construction of the proposed item-attribute graph $\mathcal{G}$ then becomes straightforward - we treat a set of all the old items as well as all their acquired attributes as nodes and then adding non-directed edges between each item and its attributes. I.e., the node set of $\mathcal{G}$ is $\mathcal{V} = \mathcal{I}_{old}\cup \mathcal{A}$ and the edge set of $\mathcal{G}$ is $\mathcal{E} = \{(i,a)|i \in \mathcal{I}_{old}, a \in \mathcal{A}, i\text{ has attribute }a\}$. 

\subsection{Item-attribute Graph Pre-training}
\label{sec:pre-training}
Figure \ref{fig:framework} (c) shows the multi-task pre-training of the proposed item-attribute graph. The pre-training process aims to fuse knowledge from the various available data sources, i.e., item contents $\mathcal{C}$, historical rating scores, and review texts of the old items $\mathcal{Y}_{old}$, into the item-attribute graph. To facilitate knowledge transfer, we design specific submodules concerning the natural forms of the data sources as well as pre-training tasks with unified losses that consistently adhere to the property of
alignment-and-uniformity \cite{wang2020understanding}, as introduced in Section \ref{sec:task1} - \ref{sec:task3}. 
The pre-training is then conducted in an MTL manner, as detailed in Section \ref{sec:mtl}.
\subsubsection{Task 1: Item-attribute Correlation Modeling}
\label{sec:task1}
The first pre-training task integrates the natural language semantics of the attributes and the higher-order item-attribute correlations into the item-attribute graph (more specifically, into the representations of its nodes). 
We adopt PLMs, \textit{e.g.}, BERT \cite{devlin2018bert}, and GNNs, \textit{e.g.}, GCN \cite{kipf2016semi} for knowledge extraction and then design alignment-and-uniformity-based loss terms for knowledge transferring.

As shown in Figure \ref{fig:framework} (c), we input the attributes and the items (each item is represented by the concatenation of its attributes) into PLM to get their natural language embeddings. We input these embeddings into an interpreter, i.e., a multi-layer perceptron (MLP), to filter out any irrelevant information and map them into a lower dimensional space for easier downstream modeling.
We then propagate the mapped embeddings over the item-attribute graph via a GNN to learn higher-order item-attribute correlations. The forward propagation follows:
\begin{equation}
\label{eq:forward_task1}
\begin{aligned}
h_1(x) = \mathrm{GNN_{1}}(\mathrm{Interpreter}(\mathrm{PLM}(x))), x \in \mathcal{I}_{old} \cup \mathcal{A}.
\end{aligned}
\end{equation}
The RHS of the above equation, referred to as the encoder of task 1, encodes the item-attribute graph. It serves as an implicit, extendable embedding matrix and is also shared either entirely or partially by the other pre-training tasks (discussed in the following sections).

Instead of adopting InfoNCE \cite{gutmann2010noise} or BPR \cite{rendle2012bpr} losses that are commonly used in Graph Pre-training \cite{hu2020gpt}, we design the first pre-training task with self-supervised loss that adhere to the alignment and uniformity properties \cite{wang2020understanding}, which are favorable in contrastive learning and essential for alleviating the sparsity issue. Specifically, task 1 minimizes $\mathcal{L}_1$ as:
\begin{equation}
\label{eq:loss_task1}
\mathcal{L}_1 = \mathcal{L}_{a1}+\lambda\mathcal{L}_{u1}, 
\end{equation}
\begin{equation}
\mathcal{L}_{a1} = \displaystyle \mathop{\mathbb{E}}_{(i,a)\sim p_{\mathrm{pos\_ia}}}\parallel h_1(i) - h_1(a)\parallel^2,
\end{equation}
\begin{equation}
\label{eq:uniformity}
\begin{aligned}
\mathcal{L}_{u1} = \displaystyle \mathrm{log}\mathop{\mathbb{E}}_{(i,i')\sim p_{\mathrm{item}}}e^{-2\parallel h_1(i) - h_1(i')\parallel^2}/2+
\mathrm{log}\mathop{\mathbb{E}}_{(a,a')\sim p_{\mathrm{attr}}}e^{-2\parallel h_1(a) - h_1(a')\parallel^2}/2.
\end{aligned}
\end{equation}
$\mathcal{L}_{a1}$ and $\mathcal{L}_{u1}$ are alignment and uniformity loss terms, respectively. $p_{\mathrm{pos\_ia}}$ denotes the distribution of the positive item-attribute pairs. $p_{\mathrm{item}}$ and $p_{\mathrm{attr}}$ denote the data distributions of the items and the attributes. In practice, we use in-batch $(i,a)$, $(i,i')$, and $(a,a')$ pairs as these instances are consistent with the actual data distributions, i.e., $p_{\mathrm{pos\_ia}}$, $p_{\mathrm{item}}$, and $p_{\mathrm{attr}}$.  
Intuitively, $\mathcal{L}_{a1}$ aligns the features of each item with the features of its attributes to capture their correlations while $\mathcal{L}_{u1}$ highlights such correlations by pushing each item/attribute away from the remaining items/attributes.

\subsubsection{Task 2: Item-item Correlation Modeling}
\label{sec:task2}
Pre-training task 2 seeks to fuse the correlations between the old items into the items-attribute graph. The users’ historical purchase sequences can reflect the correlations between these items, i.e., two items are considered correlated if they were purchased by the same user. Following this intuition, we leverage a Transformer-based sequential encoder to model the historical rating sequences, as shown in Figure \ref{fig:framework} (c). The architecture of our sequential encoder is similar to \cite{sun2019bert4rec} (we choose its bidirectional self-attention over the unidirectional self-attention adopted by other sequential encoders \cite{kang2018self}, as the former is recently shown to perform better with proper configuration \cite{petrov2022systematic}). Different from \cite{sun2019bert4rec}, which randomly initializes the item embeddings, we use the encoder of task 1 (Section \ref{sec:task1}) to serve as an implicit and extendable embedding matrix. The forward propagation follows:
\begin{equation}
\label{eq:forward_task2}
h_2(u) = \text{FFN}(\text{BiAttn}(\text{mask}(h_1(u)) + p(u))),
\end{equation}
where $u$ denotes a user's historical purchase sequence, which is a sequence of items. $h_1(u)$ is a sequence of initial embeddings of the items in $u$, acquired using the encoder of task 1 (Equation \ref{eq:forward_task1}). $\text{mask}(\cdot)$ randomly masks out some items in $u$, i.e., replace their embeddings with an embedding that represents a special token [mask]. $p(u)$ encodes the positional information of the items in $u$. BiAttn and FFN denote the stacks of multi-head self-attention layers and point-wise feed-forward layers, which are the same as in BERT4Rec. $h_2(u)$ is a sequence of embeddings of the items in $u$, enhanced with sequence-level item-item correlations.

Pre-training task 2 is then designed to predict the masked items given the remaining contextual items in the sequences. Instead of leveraging the cross-entropy loss, which is widely adopted by the existing sequential encoders including BERT4Rec, we design loss terms to adhere to the alignment and uniformity properties \cite{wang2020understanding}. These loss terms are consistent with the losses of the other pre-training tasks (Section \ref{sec:task1} and \ref{sec:task3}), which help the tasks to cooperate with each other in multi-task pre-training (illustrated in Section \ref{sec:mtl}). The loss of task 2 is defined as follows:
\begin{equation}
\label{eq:loss_task2}
\mathcal{L}_2 = \mathcal{L}_{a2}+\lambda\mathcal{L}_{u2}, 
\end{equation}
\begin{equation}
\mathcal{L}_{a2} = \displaystyle \mathop{\mathbb{E}}_{i\sim p_{\mathrm{mask}}}\parallel h_1(i) - h_2(i)\parallel^2, \quad\mathcal{L}_{u2} = \displaystyle \mathrm{log}\mathop{\mathbb{E}}_{(i,i')\sim p_{\mathrm{item}}}e^{-2\parallel h_1(i) - h_1(i')\parallel^2}/2.
\end{equation}

$\mathcal{L}_{a2}$ is the alignment loss term. The uniformity loss term $\mathcal{L}_{u2}$ follows Equation \ref{eq:uniformity} but with only the first term as the attributes are not involved in this task. $p_{\mathrm{mask}}$ denotes the distribution of the masked items and is approximated by using the in-batch instances. Intuitively, $\mathcal{L}_{a2}$ aligns $h_2(i)$, which encodes the item-item correlations underneath the purchase sequences, with $h_1(i)$, which are the item embeddings in the item-attribute graph. In this way, the item-item correlations are fused into the item-attribute graph.

\subsubsection{Task 3: Item-review Correlation Modeling}
\label{sec:task3}
Review texts contain important subjective observations made by the users and describe the subtle aspects of the items. However, review texts are ignored by the existing cold-start recommenders \cite{wu2021towards, hou2022towards} as they are high in volume, long, and noisy, and thus are difficult to explore. Our pre-training task 3 seeks to transfer knowledge from the review texts of the existing items in an efficient and effective manner. 
Specifically, as shown in Figure \ref{fig:framework} (a), we pre-process the review texts to extract phrases with sentiments. We first leverage phrase-level sentiment analysis \cite{zhang2014users} tool kits  \footnote{\url{https://github.com/evison/Sentires}}\textsuperscript{,}\footnote{\url{https://github.com/lileipisces/Sentires-Guide}} and get a set of (noun phrase, opinion word, sentiment score) tuples (the set is formally referred to as context-dependent sentiment lexicon). Concerning the huge size of the set caused by the diverse opinion words, we further filter the tuples and only keep the sentiments and the noun phrases. E.g., (lock, easy to open, 1) is mapped into `good lock' and (hinge, broken, -1) is mapped into `bad hinge'. We refer to the results as the \textit{review terms}, to be distinguished from the item attributes. 
Review terms that are too common or too rare are filtered out to simplify the graph and avoid introducing less informative item correlations. The remaining review terms, along with their correlated items, form a bipartite item-review term graph in a manner similar to how the item-attribute graph is constructed (Section \ref{sec:construction}). The items and review terms are embedded with the PLM and Interpreter of task 1 and then propagated over the item-review term graph via GNN layers. The forward propagation of task 3, as shown in Figure \ref{fig:framework} (c), follows:
\begin{equation}
\label{eq:forward_task3}
\begin{aligned}
h_3(x) = \mathrm{GNN_{3}}(\mathrm{Interpreter}(\mathrm{PLM}(x))), x \in \mathcal{I}_{old} \cup \mathcal{R},
\end{aligned}
\end{equation}
where $\mathcal{R}$ denotes a set of all the review terms. $\mathrm{GNN_{3}}$ does not share parameters with $\mathrm{GNN_{1}}$ since it aims to model the higher-order item-review term correlations instead of the item-attribute correlations. Note how the encoders of task 3 and task 1 partially overlap, by using the same PLM and Interpreter, to enable knowledge sharing. 

The loss of task 3 follows the forms of the losses of the previous two tasks and is defined as:
\begin{equation}
\label{eq:loss_task3}
\mathcal{L}_3 = \mathcal{L}_{a3}+\lambda\mathcal{L}_{u3}, 
\end{equation}
\begin{equation}
\mathcal{L}_{a3} = \displaystyle \mathop{\mathbb{E}}_{(i,r)\sim p_{\mathrm{pos\_ir}}}\parallel h_3(i) - h_3(r)\parallel^2,
\end{equation}
\begin{equation}
\label{eq:uniformity_task3}
\begin{aligned}
\mathcal{L}_{u3} = \displaystyle \mathrm{log}\mathop{\mathbb{E}}_{(i,i')\sim p_{\mathrm{item}}}e^{-2\parallel h_3(i) - h_3(i')\parallel^2}/2+
\mathrm{log}\mathop{\mathbb{E}}_{(r,r')\sim p_{\mathrm{review}}}e^{-2\parallel h_3(r) - h_3(r')\parallel^2}/2.
\end{aligned}
\end{equation}
$\mathcal{L}_{a3}$ is the alignment loss term. The uniformity loss term $\mathcal{L}_{u3}$ follows Equation \ref{eq:uniformity} but its second term contrasts review terms other than attributes. 
$p_{\mathrm{review}}$ is the distribution of the review terms and is approximated with in-batch $(r,r')$ pairs.
$p_{\mathrm{pos\_ir}}$ is the distribution of the positive item-review term pairs and is approximated by using the in-batch instances. Intuitively, $\mathcal{L}_{a3}$ aligns the features of each item with the features of its review terms to capture their correlations. Such correlations are fused into the item-attribute graph through the shared parameters of the encoders of task 3 and task 1.
\subsubsection{Multi-task Item-attribute Graph Pre-training}
\label{sec:mtl}
To pre-train a knowledgeable item-attribute graph,~\framework~leverages MTL and conducts the pre-training tasks introduced in Sections \ref{sec:task1} - \ref{sec:task3} concurrently. The encoder of task 1, which encodes the item-attribute graph, is shared by task 2 and partially by task 3, thus enabling a fusion of the knowledge learned in all three pre-training tasks into the item-attribute graph.

It is essential, however, for the pre-training tasks to cooperate with each other and facilitate positive transfer among themselves. As shown in Sections \ref{sec:task1} - \ref{sec:task3}, we design unified losses for the pre-training tasks. 
Despite the extensive studies on contrastive learning \cite{you2023rethinking, you2022mine, you2023actionplus, you2023bootstrapping}, our design stands out for its novelty as it aligns and contrasts various embedding spaces within a MTL scenario.
Our losses help to fuse knowledge from the various data sources into the item-attribute graph and mitigate the negative transfer between the tasks observed when adopting heterogeneous task losses (empirically shown in Section \ref{sec:ablation_study}). To further facilitate the positive transfer, we adopt unitary scalarization, i.e., simply minimizes the weighted sum of the task losses as it is shown, by recent studies in MTL \cite{kurin2022defense, xin2022current}, to outperform ad-hoc multi-task optimization algorithms when hyperparameters such as the task weights, the regularization term weight, and the learning rate, are properly tuned (we carefully tune the hyperparameters in Section \ref{sec:ablation_study}). The overall training objective is:
\begin{equation}\mathcal{L} =\sum_{i=1}^K \boldsymbol{w_i}\mathcal{L}_i,
\label{eq:overall_loss}
\end{equation}
where $K=3$ denotes the number of pre-training tasks. $\boldsymbol{w}>0$ are the weights of the tasks.

\subsection{SCS Item Embedding with Pre-trained Item-attribute Graph}
\label{sec:inference}
Given the pre-trained item-attribute graph, the embeddings of the SCS items can be easily acquired upon the arrival of these items and rely only on their item contents, as shown in Figure \ref{fig:framework} (d). Specifically, we first extract the attributes of the SCS items from their item contents. The SCS items $\mathcal{I}_{SCS}$, along with new item attributes that are unseen during the pre-training (denoted as $\mathcal{A}_{SCS}$) are then inserted into the pre-trained item-attribute graph. Formally, the node set of the item-attribute graph, after the update, is $\mathcal{V} = \mathcal{I}_{old} \cup \mathcal{I}_{SCS} \cup \mathcal{A} \cup \mathcal{A}_{SCS}$ while the edge set after the update is $\mathcal{E} = \{(i,a)|i \in \mathcal{I}_{old} \cup \mathcal{I}_{SCS}, a \in \mathcal{A} \cup \mathcal{A}_{SCS}, i\text{ has attribute }a\}$. We then conduct forward propagation over the updated item-attribute graph using the pre-trained encoder of task 1, shown in Equation \ref{eq:forward_task1}, to get the embeddings of all the items and attributes, including the SCS ones.
The learned item embeddings thus fuse the item-attribute, item-item, and item-review term correlations captured during the multi-task pre-training.
Note that this process does not involve any fine-tuning as the SCS scenario assumes no interactions of the SCS items are available for fine-tuning.

The acquired embeddings provide a solution for the SCS item recommendation task. Specifically, a prediction on a user's preference for an SCS item can be made by calculating the dot product of their embeddings (note that using the dot productions between the items' and users' embeddings for preference estimation is commonly seen in recommendation studies \cite{hou2022towards}). The users' embeddings, on the other hand, can be calculated by averaging the embeddings of the items that are purchased by them. Note that compared to cold-start recommenders that require large amounts of sampling \cite{wu2021towards} and querying \cite{geng2022recommendation} during inference, our approach requires simple updates and propagation over the item-attribute graph, thus is much more efficient, as shown in Section \ref{sec:main_results}.


\section{Experiment}
\label{sec:experiment}
We evaluate the proposed~\framework~in the SCS setting. Section \ref{sec:experimental_setup} reports the experimental setup. Section \ref{sec:main_results} compares~\framework~ to various baselines for top-K SCS item recommendation. Section \ref{sec:ablation_study} verifies the effectiveness of~\framework's pre-training tasks and studies the effects of changing losses as well as hyperparameters. Section \ref{sec:visualization} visualizes our proposed item-attribute graph and qualitatively shows how attributes help bridge the items.

\subsection{SCS Experimental Setup}

\label{sec:experimental_setup}
\subsubsection{SCS Datasets}

\begin{table*}
\footnotesize
\centering
\begin{tabular}{l|l|llll|llll|llll}
\hline
SCS Dataset & \#Users & \#Items & Train & Val & Test & \#Attrs & Train & Val & Test & \#Ratings & Train & Val & Test \\
\hline
Yelp-SCS & 4,901 & 2,639 & 2,258 & 118 & 263 & 2,087 & 2,035 & 9 & 43 & 193,320 & 185,097 & 2,731 & 5,492 \\
Home-SCS & 96,420 & 28,672 & 24,515 & 1,290 & 2,867 & 24,879 & 24,877 & 1 & 1 & 1,343,374 & 1,277,815 & 21,454 & 44,105 \\
Sports-SCS-1 & 81,778 & 28,316 & 26,483 & 1,393 & 440 & 12,154 & 12,130 & 0 & 24 & 464,335 & 456,050 & 6,745 & 1,540 \\
Sports-SCS-2 & 81,778 & 28,332 & 26,483 & 1,393 & 456 & 12,154 & 12,130 & 0 & 24 & 464,483 & 456,050 & 6,745 & 1,688 \\
Sports-SCS-3 & 81,778 & 28,344 & 26,483 & 1,393 & 468 & 12,155 &12,130 & 0 & 25 & 464,376 & 456,050 & 6,745 & 1,581 \\
\hline
\end{tabular}
\caption{\label{dataset}
SCS Dataset statistics.}
\vspace{-3em}
\end{table*}

We experiment on three public real-world datasets, i.e., Yelp\footnote{\url{https://www.yelp.com/dataset}}, Amazon-home, and Amazon-sports\footnote{\url{https://nijianmo.github.io/amazon/\#subsets}}, which are commonly adopted in previous recommendation studies \cite{wu2021towards, geng2022recommendation}. To guarantee the SCS setting, we carefully process these datasets. 

For Yelp and Amazon-home, we first follow previous works \cite{wu2021towards} and filter out users and items with too few (<20 for Yelp and <15 for Amazon-home) interactions. We further filter out items with insufficient item contents, as recommenders rely on item contents to make SCS item recommendations. Specifically, items that have less than five attributes are filtered out. Next, we sort the items based on the lengths of their historical purchase records. After sorting, we split the items by 9:1 for training and testing, so that the warm items are used for training. We leave out the last 5\% of the training data as a validation set for hyper-parameter tuning. Finally, interactions that involve users unseen in the training set are removed from the validation and test sets.

The Amazon-sports dataset is processed differently with the aim of comparing~\framework~to the pre-trained P5 model \cite{geng2022recommendation}. P5 is pre-trained on three large datasets, including the old version of the Amazon-sports dataset \footnote{\url{https://jmcauley.ucsd.edu/data/amazon/}} and only able to predict the preferences of users seen during its pre-training. Given this, we explore the new version of the Amazon-sports dataset, which is partially overlapped with the old version, for test set construction. Specifically, we first randomly sample a set of items that are in the new version while not in the old version of Amazon-sports. These are considered candidate SCS items. We then extract the interactions of these items from the new version and remove the interactions that involve users that do not exist in the old version. The remaining interactions form a test set, in which all users are seen, and all items are unseen by the pre-trained P5 model. In this manner, we sample three test sets, each containing $\sim$450 SCS items. The training and validation sets are then constructed from the old version of Amazon-sports in the same manner as Yelp and Amazon-home datasets.

We denote the processed datasets as Yelp-SCS, Home-SCS, and Sports-SCS-1/2/3, respectively. Table \ref{dataset} summarizes the statistics of these datasets. Note that for each processed dataset, the SCS setting is guaranteed as the interactions that involve the validation and test items are completely removed from the training set.


\subsubsection{SCS Models}
We compare~\framework~to content-based methods, Cold-start recommenders, and pre-training-based methods that accommodate SCS items.
Content-based baselines are: 1) \textbf{BERT} \cite{devlin2018bert}. As a PLM, it enables a direct acquisition of the embeddings of the items by taking the concatenations of their attributes as input. 2) \textbf{BERT+R}. To study the effects of leveraging the review texts in naive manners, for each old item, we further get a set of the BERT-based embeddings of all its reviews, then calculate the average of the attribute-based embedding and the review embeddings for the item's final embedding. 3) \textbf{AFM} \cite{xiao2017attentional}, which is a variant of FM and leverages the attention mechanism to model the interactions of the input features.
For Cold-start recommenders, we consider: 4) \textbf{DropoutNet} \cite{volkovs2017dropoutnet}, which adopts a two-tower neural network and applies dropout to input mini-batches to reconstruct preference embeddings from content embeddings. 5) \textbf{Heater} \cite{zhu2020recommendation}, which adopts a randomized training mechanism for better reconstruction. For pre-training-based methods, we consider: 6) \textbf{IDCF-HY} \cite{wu2021towards}, which is the SCS extension of the graph- and meta-learning-based IDCF \cite{wu2021towards} and relies on contents other than interactions for adaptation. 7) \textbf{GPT-GNN} \cite{hu2020gpt}, which is a generative graph pre-training method. 8) \textbf{UniSRec} \cite{hou2022towards}, which embeds items with PLM and extends sequential recommenders \cite{kang2018self, liu2021augmenting} to the inductive setting. 9) \textbf{P5} \cite{geng2022recommendation}, which adopts a PLM backbone, i.e., T5 \cite{raffel2020exploring} and further pre-trains it by distorting recommendation data into natural language sequences. 
Furthermore, we compare to 10) the pre-trained, cross-domain version of UniSRec, denoted as \textbf{UniSRec-PT}. UniSRec-PT leverages knowledge transferred from other domains (five Amazon datasets, i.e., Food, CDs, Kindle, Movies, and Home, with 14,029,229 interactions in total). Note that UniSRec-PT is not considered as a baseline since it is pre-trained on large datasets (75-220 times larger than the pre-training data of~\framework) from different domains while~\framework~ only leverages data from the target domain. 

\subsubsection{SCS Experiment Setting}
\label{sec:SCS_experiment_setting}
For~\framework, we set the batch size to 512, the embedding dimension to 64, the learning rate to 0.005, the pre-training tasks' weights $\boldsymbol{w}$ to [0.6, 0.2, 0.2], and the regularization term weight $\lambda$ to 0.6, the PLM to SBERT \cite{reimers2019sentence} (the effects of changing the hyperparameters are observed in Section \ref{sec:ablation_study}). We set the number of convolutional layers to 1 for $\text{GNN}_1$ and $\text{GNN}_3$. For the BERT4Rec component in the submodule for our pre-training task 2, we set the maximum sequence length to 100, the number of self-attention layers to 1, the number of attention heads to 1, and the item masking probability to 0.2.
When the BERT4Rec component is not involved (as in some of the ablation studies in \ref{sec:ablation_study}), we adopt early stopping with a patience of 50 epochs using the pre-training loss (Equation \ref{eq:overall_loss}) as an indicator. Otherwise, we follow \cite{petrov2022systematic} and adopt early stopping with a patience of 200 epochs.
Consider the large volume of the review texts, we randomly sample 100 reviews per item to construct the item-review term graph for our pre-training task 3.
As mentioned in Section \ref{sec:inference}, we calculate the users' embeddings by averaging the embeddings of their
purchased items. We then predict a user’s preference for an item by calculating the dot product of their embeddings.
For AFM, we adopt user embedding, item embedding, and rating score as the input features.
For DropoutNet and Heater, we pre-train a Matric Factorization (MF) model on the training data to provide the required user/item collaborative embedding. The model structure and hyperparameter setting are consistent with the original paper.
For IDCF-HY, UnisRec, and GPT-GNN, we adopt the experiment settings as reported in their original papers. For UnisRec, we acquire a user's embedding from its universal sequence representation module, and a test item's embedding from its MoE-enhanced adaptor then calculate the dot product of these embeddings for the predicted rating.
For P5, we adopt the pre-trained version \footnote{\url{https://huggingface.co/makitanikaze/P5_sports_base}}. Note we only report P5's performance on the Sports-SCS datasets as the pre-trained P5 model for the other two datasets is unavailable. We query P5 with the `Z-3' template. One query predicts only one interaction, therefore, we query all the test items for each user before ranking. For BERT and BERT+R, we adopt Hugging Face pre-trained models, i.e., `bert-large-cased'. The prediction process is consistent with~\framework~after the acquisition of the item embeddings.
We use a 12-core Intel Xeon CPU E5-2620 v3@2.40GHz with 64GB RAM and 3×NVIDIA GeForce GTX 1080 Ti GPU and report the mean over 5 runs for all experiments. 

\subsubsection{Evaluation Metrics}
Following the previous recommendation studies \cite{hou2022towards, wu2021towards}, we adopt two widely used metrics, i.e., NDCG and Recall. We report Recall@N
and NDCG@N, where N is set to 5, 20, and 40.
\subsection{SCS Item Recommendation}
\label{sec:main_results}

\begin{table*}
\centering
\footnotesize
\begin{tabular}{p{0.69cm}p{0.99cm}|ccccccccc|P{0.8cm}|c}
\hline
Dataset & Metric & BERT & BERT+R & AFM & DropoutNet & Heater & IDCF-HY & GPT-GNN & P5 & UniSRec & \framework~ & $\Delta$ (\%) \\
\hline
 & NDCG@5 & 0.0125 & 0.0128 & 0.0132 & 0.0127 & 0.0128 & 0.0113 & 0.0032 & / & \underline{0.0161} & \textbf{0.0265} & 64.60\\
 & NDCG@20 & 0.0301 & 0.0297 & 0.0308 & 0.0304 & 0.0301 & 0.0292 & 0.0091 & / & \underline{0.0383} & \textbf{0.0556} & 45.17\\
 Yelp & NDCG@40 & 0.0480 & 0.0473 & 0.0504 & 0.0488 & 0.0480 & 0.0459 & 0.0157 & / & \underline{0.0573} & \textbf{0.0863} &50.61\\
\cline{2-13}
 -SCS & Recall@5 & 0.0182 & 0.0179 & 0.0170 & 0.0172 & 0.0179 & 0.0161 & 0.0042 & / & \underline{0.0257} & \textbf{0.0371} & 44.36\\
 & Recall@20 & 0.0756 & 0.0743 & 0.0729 & 0.0734 & 0.0743 & 0.0736 & 0.0234 & / & \underline{0.0989} & \textbf{0.1282} & 29.63\\
 & Recall@40 & 0.1541 & 0.1504 & 0.1566 & 0.1495 & 0.1485 & 0.1436 & 0.0515 & / & \underline{0.1816} & \textbf{0.2354} & 29.63\\
\hline
 & NDCG@5 & 0.0010 & 0.0010 & 0.0011 & 0.0010 & 0.0011 & / & 0.0009 & / & \underline{0.0064} & \textbf{0.0095} & 48.44\\
 & NDCG@20 & 0.0027 & 0.0025 & 0.0026 & 0.0025 & 0.0026 & / & 0.0056 & / & \underline{0.0112} & \textbf{0.0166} & 48.21\\
 Home & NDCG@40 & 0.0042 & 0.0037 & 0.0041 & 0.0041 & 0.0040 & / &  0.0121 & /  & \underline{0.0148} & \textbf{0.0215} & 45.27\\
\cline{2-13}
 -SCS & Recall@5 & 0.0017 & 0.0016 & 0.0018 & 0.0016 & 0.0018 & / & 0.0019 & / & \underline{0.0094} & \textbf{0.0139} & 47.87\\
 & Recall@20 & 0.0073 & 0.0070 & 0.0070 & 0.0070 & 0.0071 & / & 0.0179 & / & \underline{0.0263} & \textbf{0.0383} & 45.63\\
 & Recall@40 & 0.0144 & 0.0135 & 0.0139 & 0.0140 & 0.0134 & / & \underline{0.0483} & / & 0.0426 & \textbf{0.0609} & 26.08\\
\hline
 & NDCG@5 & 0.0052 & 0.0051 & 0.0044 & 0.0082 & 0.0054 & 0.0049 & 0.0000 & 0.0047 & \underline{0.0303} & \textbf{0.0324} & 6.93\\
 & NDCG@20 & 0.0154 & 0.0154 & 0.0135 & 0.0190 & 0.0152 & 0.0147 & 0.0000 & 0.0182 & \underline{0.0613} & \textbf{0.0663} & 8.16\\
 Sports & NDCG@40 & 0.0264 & 0.0259 & 0.0283 & 0.0274 & 0.0244 & 0.0252 & 0.0004 & 0.0231 & \underline{0.0830} & \textbf{0.0884} & 6.51\\
\cline{2-13}
 -SCS-1 & Recall@5 & 0.0075 & 0.0076 & 0.0092 & 0.0124 & 0.0097 & 0.0072 & 0.0000 & 0.0080 & \underline{0.0528} & \textbf{0.0571} & 8.14\\
 & Recall@20 & 0.0448 & 0.0446 & 0.0411 & 0.0505 & 0.0446 & 0.0446 & 0.0000 & 0.0473 & \underline{0.1624} & \textbf{0.1760} & 8.37\\
 & Recall@40 & 0.0981 & 0.0970 & 0.1012 & 0.0910 & 0.0889 & 0.0971 & 0.0021 & 0.0744 & \underline{0.2666} & \textbf{0.2831} & 6.19\\
\hline
 & NDCG@5 & 0.0043 & 0.0046 & 0.0050 & 0.0051 & 0.0067 & 0.0039 & 0.0002 & 0.0064 & \underline{0.0271} & \textbf{0.0297} & 9.59\\
 & NDCG@20 & 0.0121 & 0.0113 & 0.0168 & 0.0114 & 0.0158 & 0.0107 & 0.0022 & 0.0148 & \underline{0.0636} & \textbf{0.0660} & 3.77\\
 Sports & NDCG@40 & 0.0203 & 0.0258 & 0.0283 & 0.0196 & 0.0269 & 0.0185 & 0.0037 & 0.0232 & \underline{0.0823} & \textbf{0.0904} & 9.84\\
\cline{2-13}
 -SCS-2 & Recall@5 & 0.0073 & 0.0071 & 0.0092 & 0.0069 & 0.0096 & 0.0068 & 0.0003 & 0.0098 & \underline{0.0455} & \textbf{0.0504} & 10.77\\
 & Recall@20 & 0.0347 & 0.0329 & 0.0497 & 0.0303 & 0.0421 & 0.0311 & 0.0075 & 0.0386 & \underline{0.1621} & \textbf{0.1788} & 10.30\\
 & Recall@40 & 0.0747 & 0.0706 & 0.1062 & 0.0701 & 0.0964 & 0.0688 & 0.0151 & 0.0822 & \underline{0.2528} & \textbf{0.2974} & 17.64\\
\hline
 & NDCG@5 & 0.0055 & 0.0051 & 0.0032 & 0.0032 & 0.0059 & 0.0050 & 0.0000 & 0.0056 & \underline{0.0358} & \textbf{0.0359} & 0.28\\
 & NDCG@20 & 0.0117 & 0.0112 & 0.0107 & 0.0108 & 0.0113 & 0.0110 & 0.0002 & 0.0138 & \underline{0.0628} & \textbf{0.0692} & 10.19\\
 Sports & NDCG@40 & 0.0210 & 0.0195 & 0.0197 & 0.0209 & 0.0224 & 0.0192 & 0.0005 & 0.0229 & \underline{0.0826} & \textbf{0.0924} & 11.86\\
\cline{2-13}
 -SCS-3 & Recall@5 & 0.0073 & 0.0068 & 0.0056 & 0.0066 & 0.0102 & 0.0066 & 0.0000 & 0.0086 & \underline{0.0565} & \textbf{0.0569} & 0.71\\
 & Recall@20 & 0.0308 & 0.0298 & 0.0336 & 0.0345 & 0.0289 & 0.0268 & 0.0007 & 0.0367 & \underline{0.1515} & \textbf{0.1758} & 16.04\\
 & Recall@40 & 0.0763 & 0.0741 & 0.0800 & 0.0836 & 0.0841 & 0.0705 & 0.0021 & 0.0777 & \underline{0.2486} & \textbf{0.2892} & 16.33\\
\cline{2-13}
 & Pred. time & $<60$ s & $<60$ s & $<60$ s & $<60$ s & $<60$ s & 11 hrs & $<60$ s & 20 hrs & $<60$ s & $<60$ s & / \\
\hline
\end{tabular}
\caption{\label{main-result}
Top-K SCS item recommendation performance. The best results are boldfaced and the second-best ones are underlined. $\Delta$ indicates the relative improvement upon the best baseline. `/' indicates an unachievable result (e.g., the inference of IDCF-HY on Home-SCS takes >120 hrs while P5 does not recognize the Yelp-SCS and Home-SCS users, which are unseen during P5's pre-training).}
\vspace{-3em}
\end{table*}

\begin{table}
\centering
\footnotesize
\begin{tabular}{p{2cm}|P{2cm}P{2cm}P{2cm}P{2cm}}
\hline
 & Yelp-SCS & Sports-SCS-1 & Sports-SCS-2 & Sports-SCS-3 \\
\hline
\multicolumn{5}{l}{\textbf{UnisRec-PT}\;\;\;\;\;\;\;\;\;\;Pre-trained on 14,029,229 ratings from Food, CDs, Kindle, Movies, and Home domains} \\
\multicolumn{5}{l}{$\qquad\qquad\ \ \ $\;\;\;\;\;\;\;\;\;\;\;\;+ Fine-tuned on 187,828 | 62,795 | 62,795 | 62,795 ratings} \\
 NDCG@5 & \textbf{0.0301} & 0.0307 & 0.0261 & \textbf{0.0397} \\
 NDCG@20 & \textbf{0.0667} & 0.0609 & 0.0574 & 0.0672 \\
 NDCG@40 & \textbf{0.0925} & 0.0823 & 0.0805 & 0.0877 \\
\hline
 Recall@5 & \textbf{0.0430} & 0.0473 & 0.0417 & \textbf{0.0655} \\
 Recall@20 & \textbf{0.1602} & 0.1531 & 0.1516 & 0.1624 \\
 Recall@40 & \textbf{0.2723} & 0.2573 & 0.2641 & 0.2621 \\
\hline
\hline
\multicolumn{5}{l}{\textbf{\framework~(ours)} \;\;\;Pre-trained on 187,828 | 62,795 | 62,795 | 62,795 ratings} \\
 NDCG@5 & 0.0265 & \textbf{0.0324} & \textbf{0.0297} & 0.0359 \\
 NDCG@20 & 0.0556 & \textbf{0.0663} & \textbf{0.0660} & \textbf{0.0692} \\
 NDCG@40 & 0.0863 & \textbf{0.0884} & \textbf{0.0904} & \textbf{0.0924} \\
\hline
 Recall@5 & 0.0371 & \textbf{0.0571} & \textbf{0.0504} & 0.0569 \\
 Recall@20 & 0.1282 & \textbf{0.1760} & \textbf{0.1788} & \textbf{0.1758} \\
 Recall@40 & 0.2354 & \textbf{0.2831} & \textbf{0.2974} & \textbf{0.2892} \\
\hline
\end{tabular}
\caption{\label{Xdomain}
Compare~\framework~to models pre-trained on large-volume, cross-domain data. The results of~\framework~are copied from Table \ref{main-result}. The best results are boldfaced.}
\vspace{-4em}
\end{table}

Table \ref{main-result} summarizes the top-K SCS item recommendation results. The observations are: 1) The proposed~\framework~consistently outperforms the best baseline, i.e., UniSRec, by large margins. For example,~\framework~achieves up to 64\% higher NDCGs and up to 48\% higher Recalls than UnisRec. Note that unlike UnisRec, which leverages the item contents in coarse manners (concatenate the fields then truncate the result),~\framework~extracts and models fine-grained item attributes. This shows that fine-grained item-attribute correlations are essential for SCS item recommendation and the proposed item-attribute graph as well as multi-task pre-training framework effectively explore such correlations.
2) Compared to merely using the item contents, incorporating the rating score sequences of the existing items further improves performance. Specifically, models that incorporate sequential information, i.e.,~\framework~and UniSRec, outperform the other methods. This show that the item-item correlations contained in the historical rating sequences help to reveal the correlations between the existing items and the SCS ones.
3) Simple content-based methods, i.e., BERT and AFM, perform better than or on par with some cold-start and pre-training-based recommenders, i.e., DropoutNet, Heater, IDCF-HY, GPT-GNN, and P5.
This shows that these recommenders fail to further capture the correlations between the SCS items and the old items. The reason is, they explore the available data sources in insufficient or coarse manners (e.g., DropoutNet, Heater, IDCF-HY, and GPT-GNN ignore the sequential correlations between the items, while P5 distorts the rating score sequences to natural language) and thus model noisy and unreliable item correlations.
In contrast, our proposed~\framework~constructs and pre-trains a fine-grained, knowledgeable item-attribute graph and efficiently mitigates the gap between the existing and the SCS items. This again justifies the importance of incorporating more information sources and exploring the information sources in fine-grained other than naive or coarse manners. 4) Although review texts contain valuable information, leveraging the in naive manners does not help. Specifically, BERT+R performs on par with BERT. We discuss more on proper ways to leverage review texts in Section \ref{sec:ablation_study}.
5) The inference of~\framework~is efficient. For example, the inferences of IDCF-HY and P5 on Sports-SCS-3 take hours as they require large amounts of sampling and querying, respectively. In contrast,~\framework~takes < 60 seconds as it simply needs to insert the SCS items and attributes into the pre-trained item-attribute graph and then propagate. 

Table \ref{Xdomain} compares~\framework~to UnisRec-PT, which is pre-trained on 76-224 times more, cross-domain data. Despite being greatly disadvantaged by its scarce pre-training data,~\framework~outperforms UnisRec-PT on Sports-SCS-1 and Sports-SCS-2 while performing on par on Sports-SCS-3. This shows the effectiveness and data efficiency of~\framework. We recognize, on the other hand, that cross-domain knowledge transferring may further improve~\framework's performance and leave the exploration to future work (Section \ref{sec:conclusion}). 
\subsection{Ablation Study}
\label{sec:ablation_study}

\begin{table*}
\centering
\footnotesize
\begin{tabular}{c|P{2.6cm}c|cccccc}
\hline
& Factor & Value & NDCG@5 & NDCG@20 & NDCG@40 & Recall@5 & Recall@20 & Recall@40 \\
\hline
1 & \multirow{14}{*}{task weights ($\boldsymbol{w}$)} & [1.0, 0.0, 0.0] & 0.0227 & 0.0454	& 0.0721 & 0.0312 & 0.1015 & 0.2102 \\
2 & & [0.8, 0.2, 0.0] & 0.0242	&0.0546	&0.0796	&0.0342	&0.1223	&0.2341 \\
3 & & [0.8, 0.0, 0.2] & 0.0258	&0.0548	&0.0819	&0.0335	&0.1242	&\underline{0.2388} \\
4 & & [0.6, 0.4, 0.0] & 0.0250	&0.0542	&0.0805	&0.0328	&0.1241	&0.2366 \\
5 & & [0.6, 0.0, 0.4] & \underline{0.0263}	&0.0538	&0.0800	&0.0353	&0.1223	&0.2331 \\
6 & & [0.4, 0.6, 0.0] & 0.0246	&0.0503	&0.0810	&0.0328	&0.1143	&0.2345 \\
7 & & [0.4, 0.4, 0.2] & 0.0257	&0.0521	&0.0768	&0.0336	&0.1171	&0.2225 \\
8 & & [0.4, 0.2, 0.4] & 0.0251	&0.0522	&0.0797	&0.0318	&0.1166	&0.2367 \\
9 & & [0.4, 0.0, 0.6] & 0.0261	&0.0538	&0.0793	&0.0339	&0.1231	&0.2302 \\
10 & & [0.2, 0.8, 0.0] & 0.0230	&0.0458	&0.0710	&0.0311	&0.1039	&0.2101 \\
11 & & [0.2, 0.6, 0.2] & 0.0228	&0.0462	&0.0714	&0.0307	&0.1028	&0.2096 \\
12 & & [0.2, 0.4, 0.4] & 0.0255	&0.0523	&0.0784	&0.0347	&0.1192	&0.2301 \\
13 & & [0.2, 0.2, 0.6] & 0.0213	&0.0489	&0.0741	&0.0278	&0.1158	&0.2228 \\
14 & & [0.2, 0.0, 0.8] & 0.0251	&0.0481	&0.0738	&0.0335	&0.1065	&0.2166 \\
\hline
15 & task 3 & w/o sentiments & 0.0217 & 0.0510 & 0.0790 & 0.0293 & 0.1227 & 0.2340 \\
\hline
16 & \multirow{3}{*}{regularization weight ($\lambda$)} & 0.2 & 0.0225 & 0.0498 & 0.0639 & 0.0317 & 0.0991 & 0.2005 \\
17 &  & 0.4 & 0.0258 & 0.0505 & 0.0780 & \underline{0.0370} & 0.1152 & 0.2327 \\
18 &  & 0.8 & 0.0262 & \underline{0.0558} & 0.0853 & 0.0368 & 0.1208 & 0.2274 \\
\hline
19 & \multirow{3}{*}{learning rate ($lr$)} & 0.010 & 0.0221 & 0.0495 & 0.0777 & 0.0315 & 0.1194 & 0.2291 \\
20 &  & 0.008 & 0.0255 & 0.0553 & 0.0851 & 0.0362 & 0.1269 & 0.2341 \\
21 &  & 0.003 & 0.0246 & 0.0516 & 0.0812 & \textbf{0.0371} & 0.1232 & 0.2300 \\
\hline
22 & PLM & BERT & 0.0255 & \textbf{0.0559} & \underline{0.0862} & \underline{0.0370} & \underline{0.1277} & \textbf{0.2457} \\
\hline
23 & loss function & InfoNCE + CE & 0.0146 & 0.0385 & 0.0608 & 0.0215 & 0.0990 & 0.1941 \\
\hline
24 & \multicolumn{2}{c|}{Default setting (Section \ref{sec:SCS_experiment_setting})} & \textbf{0.0265}	&0.0556	&\textbf{0.0863}	&\textbf{0.0371}	&\textbf{0.1282}	&0.2354 \\
\hline
\end{tabular}
\caption{\label{ablation}
Ablation study of~\framework~on the Yelp-SCS dataset. In line 24, The results under the default setting ($\boldsymbol{w} = $[0.6, 0.2, 0.2], $\lambda = $0.6, $lr = $0.005, PLM = SBERT, loss function = Equation \ref{eq:overall_loss}) are copied from Table \ref{main-result}. The best results are boldfaced and the second-best ones are underlined.}
\vspace{-3em}
\end{table*}

\begin{figure}%
    \centering
    \subfloat[\centering NDCG@5 w.r.t. $w_2$]{{\includegraphics[width=3.5cm]{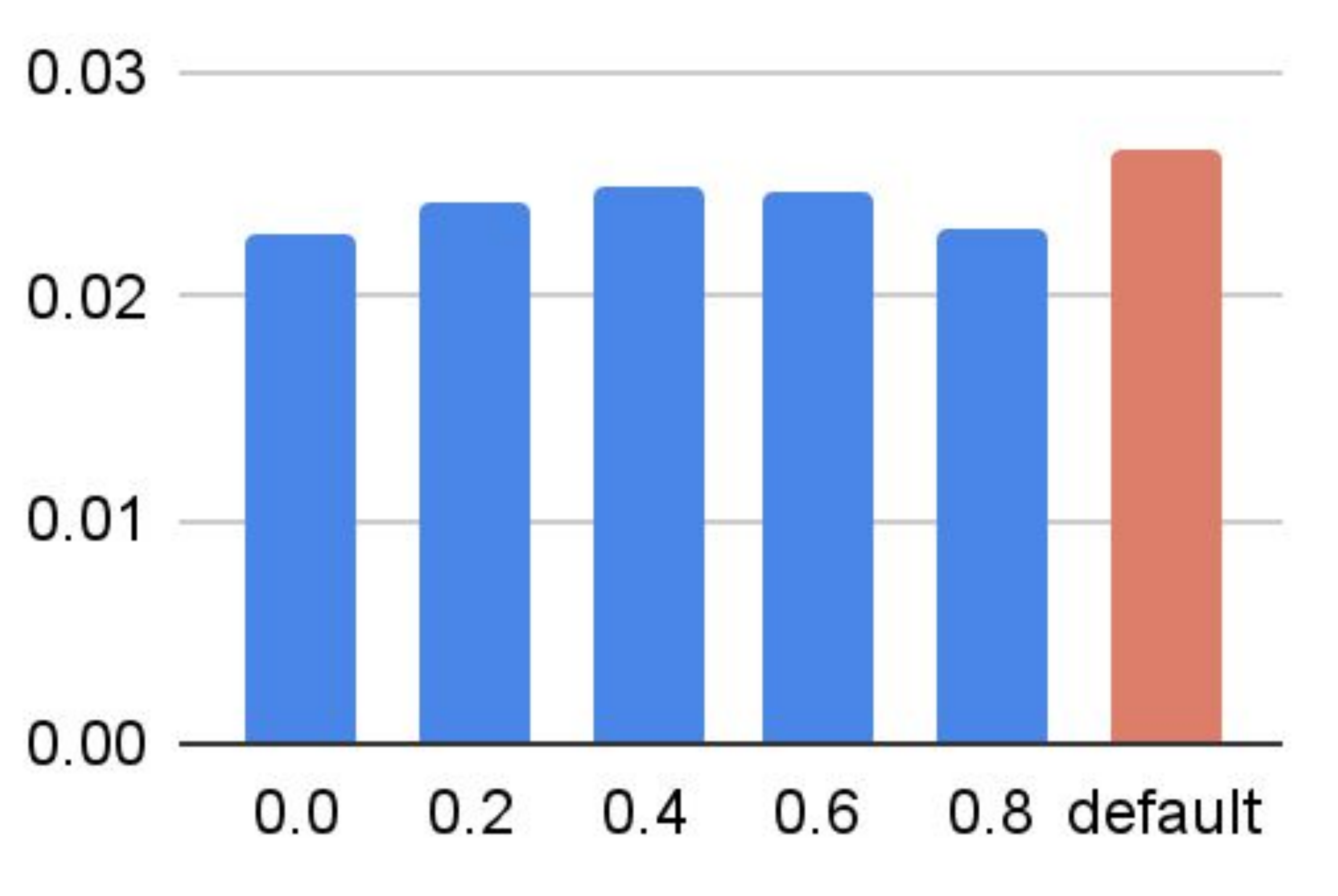} }}%
    \subfloat[\centering NDCG@5 w.r.t. $w_3$]{{\includegraphics[width=3.5cm]{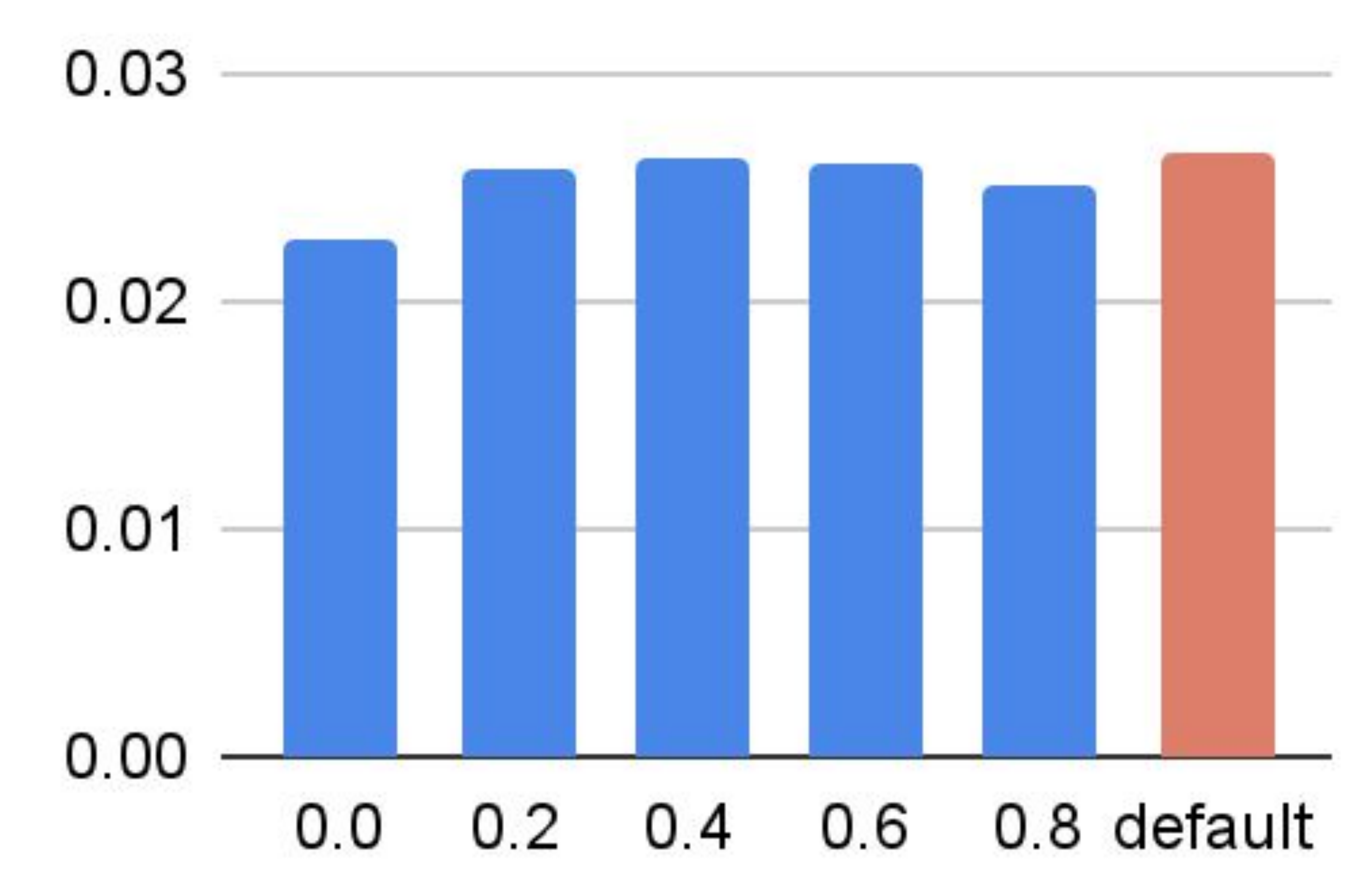} }}%
    \subfloat[\centering Loss values of lines 23]{{\includegraphics[width=3.7cm]{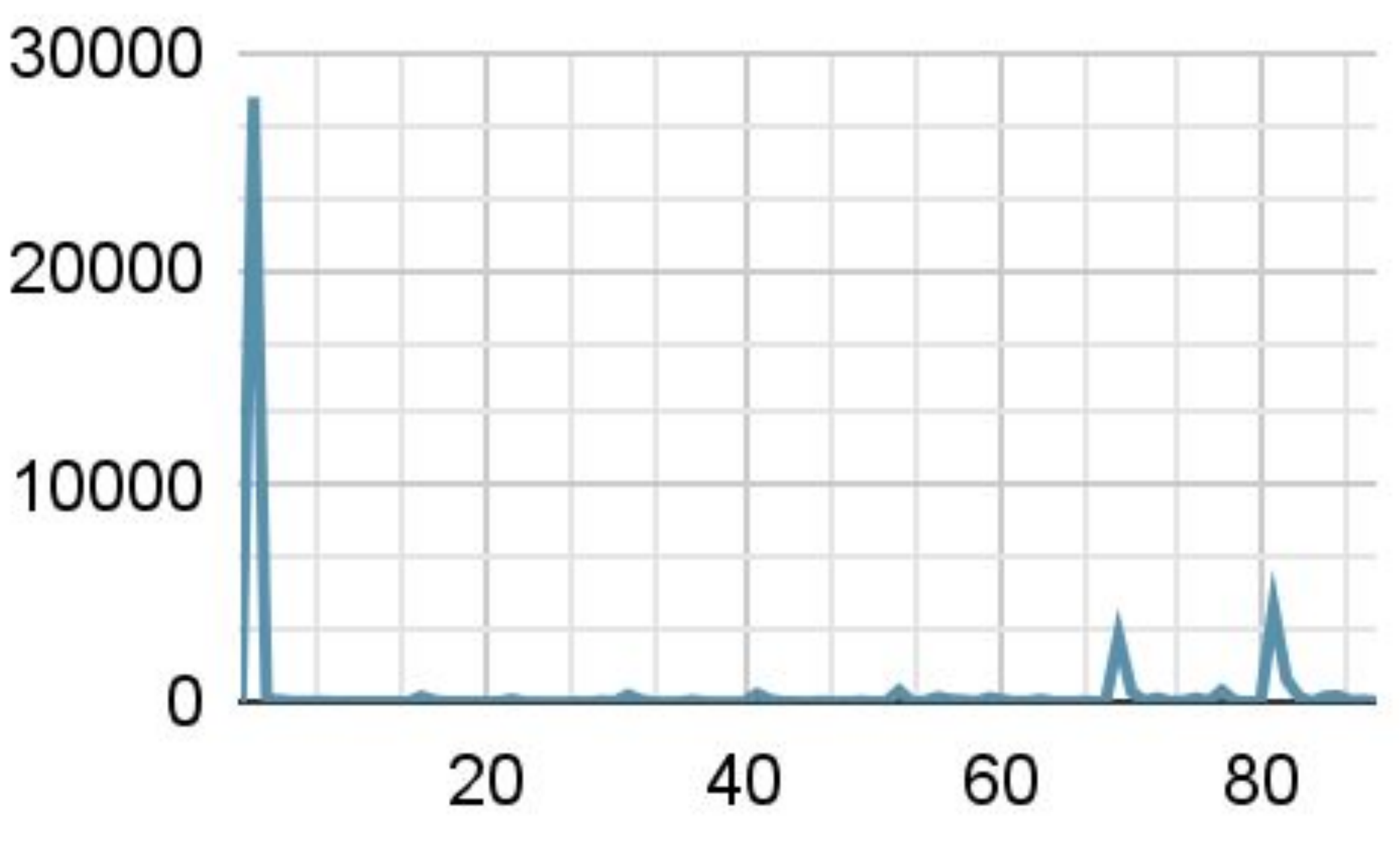} }}%
    \subfloat[\centering Loss values of lines 24 (ours)]{{\includegraphics[width=3.7cm]{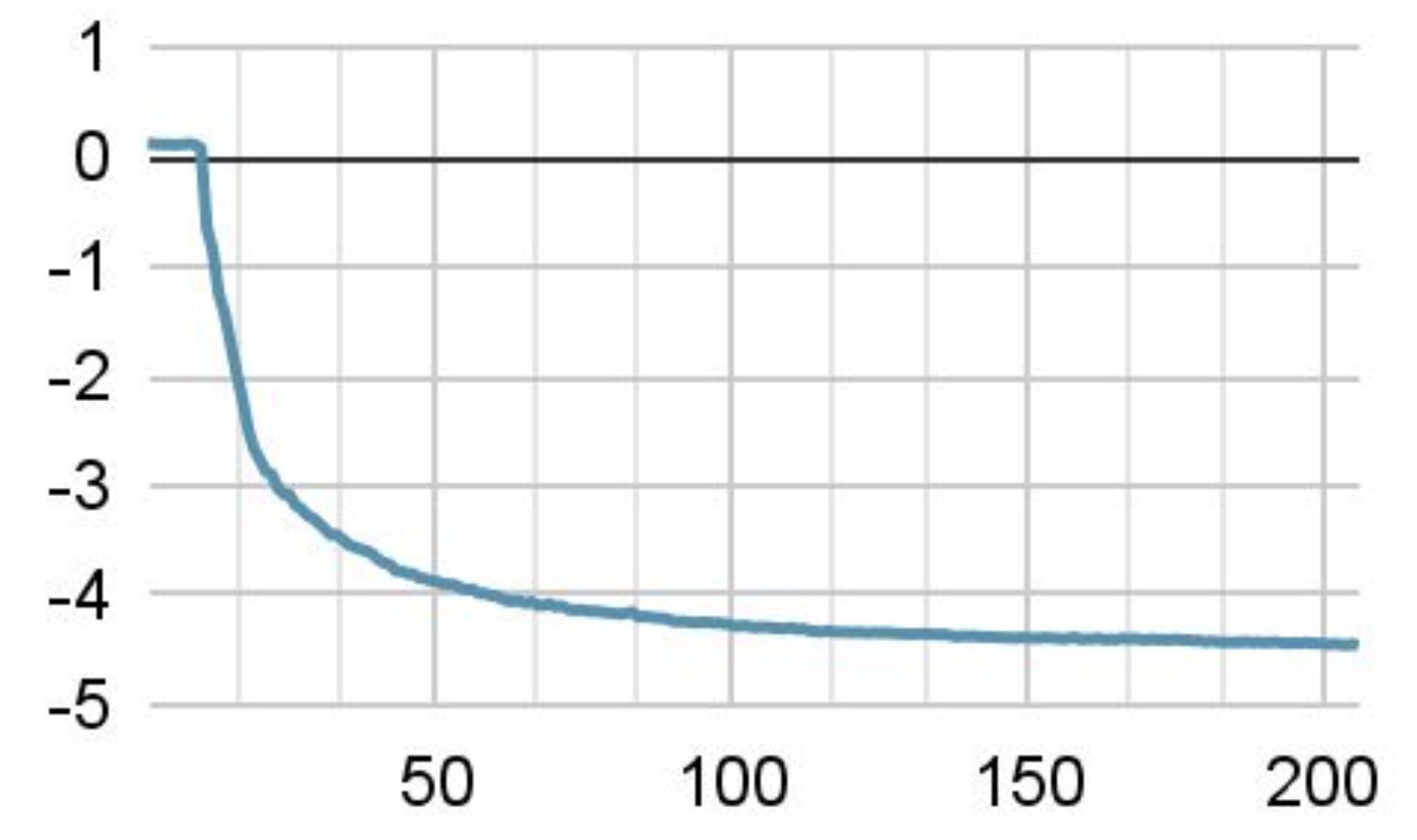} }}
    \caption{The changes in NDCG@5 when increasing the weights of task 2 (a) and task 3 (b), the loss values of lines 23 (c) and 24 (d) in Table \ref{ablation}. In (a), we fix the weight of task 3 to 0.0 and gradually increase the weight of task 2. Similarly, in (b), we gradually increase the weight of task 3. 'default' refers to the default setting (line 24 in Table \ref{ablation}). In (c) and (d), the x-axis shows the training epochs.}%
    \label{fig:loss}%
\end{figure}

We study the effects of changing the task weights $\boldsymbol{w}$, task 3 data, regularization term weight $\lambda$, learning rate (lr), PLM, and loss function of~\framework. Table \ref{ablation} and Figure \ref{fig:loss} present the results. As summarized in Figure \ref{fig:loss} (a) and (b), incorporating pre-training tasks 2 and 3 on top of task 1 helps. Comparing lines 1 - 14 and line 24 further shows that the three tasks are able to cooperate with each other during multi-task pre-training and perform the best with the weights set to [0.6, 0.2, 0.2]. In line 15, we replace the review terms in task 3 with noun phrases, i.e., we remove sentiments from the original review terms. Line 15 performs less well than line 24, indicating that the sentiments of the users help to better model item correlations and should be considered when exploring the subjective review texts. Comparing the results in lines 16 - 21 and line 24 indicates that setting $\lambda$ and lr to 0.6 and 0.005, respectively, provides the best performance. Comparing line 22 to line 24 shows that~\framework~works despite the choice of PLM and~\framework~still presents competitive results after changing its PLM from SBERT to BERT. Changing the pre-training task losses lead to a considerable drop in performance, as shown in line 23. Instead of adopting the proposed alignment-based loss terms, line 23 adopts InfoNCE loss, which is commonly used by graph pre-training methods \cite{qiu2020gcc, hu2020gpt} for tasks 1 and 3. For task 2, line 23 follows the original design of BERT4Rec \cite{sun2019bert4rec} and adopts cross-entropy loss. Line 23 also adopts Euclidean norms other than uniformity-based regularization terms. These changes cause the pre-training tasks to conflict with each other and exploding gradients are observed. As shown in Figure \ref{fig:loss} (c) and (d), the loss of line 23 fluctuates severely between 0 and 30,000, while the loss of line 12 steadily drops to around -5. This shows that the unified, alignment-and-uniformity-based pre-training task losses of~\framework~effectively facilitate the positive transfer. 
Also, note that line 23 performs on par with while the rest lines outperform the best baseline, i.e., UnisRec (Table \ref{main-result}). For example, lines 2, 4, 6, and 10 show that~\framework~outperforms UnisRec when using the same data sources (i.e., without incorporating the review texts) as the later. Line 1 further shows that~\framework~outperforms UnisRec using even less data (i.e., merely relying on the item contents). This again verifies~\framework's effectiveness.


\subsection{Item-attribute Graph Visualization}
\label{sec:visualization}
\begin{figure}%
    \centering
    \subfloat[\centering All item pairs]{{\includegraphics[width=4.6cm]{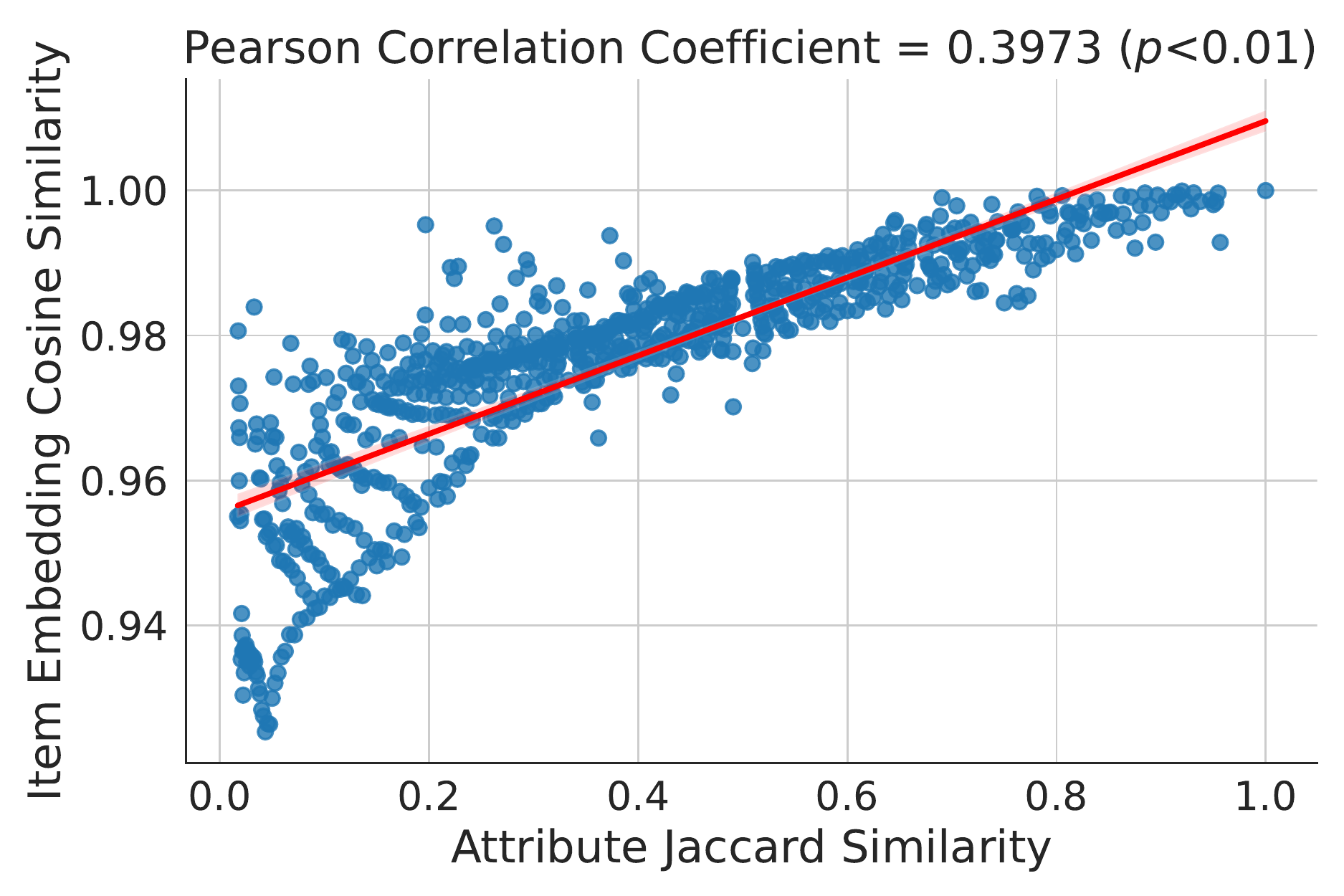} }}%
    \subfloat[\centering SCS-existing item pairs]{{\includegraphics[width=4.6cm]{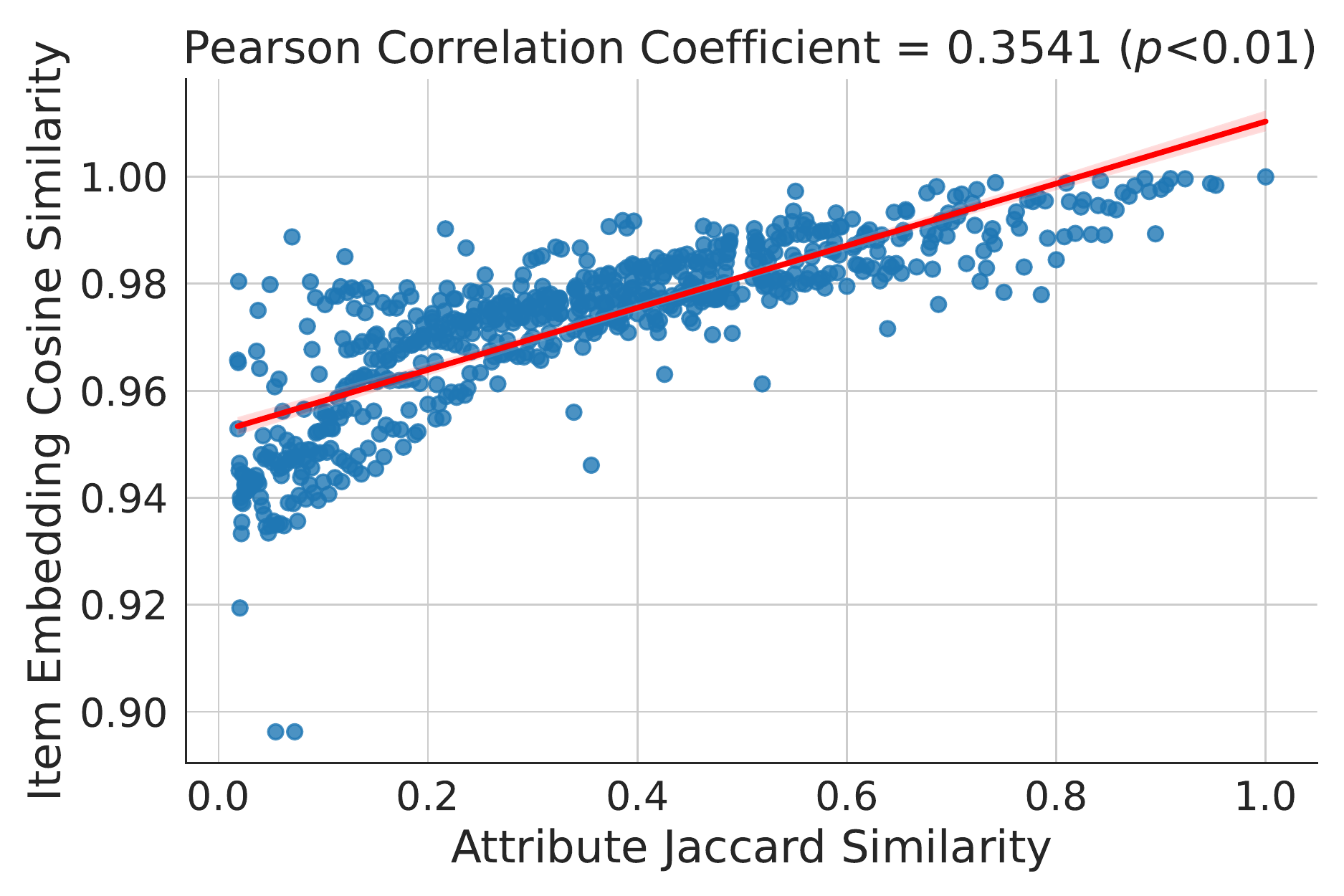} }}%
    \subfloat[\centering SCS-SCS item pairs]{{\includegraphics[width=4.6cm]{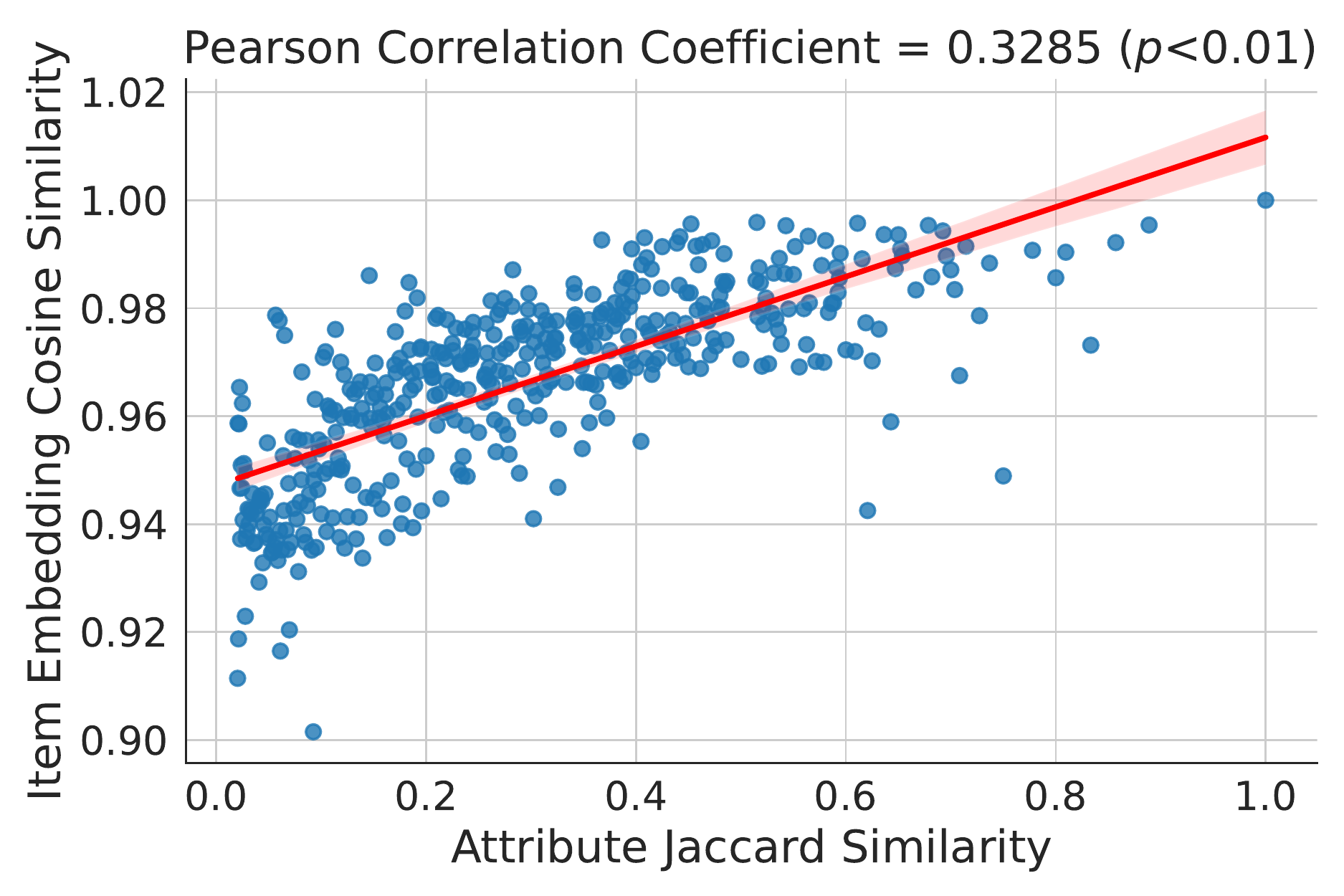} }}
    \caption{Pearson correlation coefficients between the~\framework~embedding cosine similarity and the attribute Jaccard similarity of the item pairs.}%
    \label{fig:visual}%
\vspace{-1em}
\end{figure}

We visualize the Pearson correlation coefficients between the pair-wise distance (measured by calculating the cosine similarity between the item embeddings learned via~\framework) and attributes' similarity of the item pairs, as shown in Figure \ref{fig:visual}.
We visualize all item pairs, SCS-existing item pairs, and SCS-SCS item pairs in Figure \ref{fig:visual} (a), (b) and (c), respectively. We can observe the following. 1) For all item pairs, the Pearson correlation is positive and significant at the $0.01$ level.
This shows that the item embeddings learned by~\framework~capture the correlations between the items - for the two items in each pair, the more attributes they share, the closer their embeddings are.
2) For SCS-existing item pairs, the Pearson correlation is also positive and significant at the $0.01$ level, indicating that~\framework~ effectively transfers attribute-related information to the new items and connects the new items to the old items. 
3) For SCS-SCS item pairs, the Pearson correlation is also positive and significant at the $0.01$ level. It shows even if two items are both strictly cold start items,~\framework~can generate positively correlated embeddings with respect to their shared attributes.
4) The order of Pearson coefficient between the item embedding similarity and the attribute similarity on all item pairs (0.3973) $>$ SCS-existing item pairs (0.3541) $>$ SCS-SCS pairs (0.3285). This is justifiable since all the existing items are explicitly trained to align with the attributes, which makes the Pearson coefficients on all item pairs higher. SCS items, unseen during pre-training, are not directly aligned with the attributes, which makes the coefficient on SCS-existing items pairs relatively lower. It will be even lower if two items are both unseen during the pre-training stage.



\section{Related work}

\noindent \textbf{Cold-start Recommenders.} 
Recommendation methods suffer from the out-of-distribution problem in cold-start scenarios where the well-established, dominating ID-based approaches deteriorate \cite{wheretogo2023, wu2021towards}. 
The existing cold-start recommenders mainly aim at the few-shot cold-start setting and require a few interactions of the test items for fine-tuning or adaptation. Methods based on meta-learning \cite{lu2020meta, vartak2017meta, lee2019melu, wu2021towards, pang2022pnmta, du2022metakg, hao2021pre, lin2021task, zhu2021learning, pan2019warm, dong2020mamo} are commonly seen. To address the SCS setting, in which the interactions of the test items are completely inaccessible, content-based methods such as wide\&deep \cite{cheng2016wide}, Factorization Machines (FM) \cite{rendle2010factorization} and its variants \cite{xiao2017attentional} can be applied, as they rely merely on the contents (auxiliary user-profiles and item descriptions) of the users and items for making recommendations. Specific types of cold-start recommenders are also able to work in the SCS setting. These methods typically reconstruct the preference (Collaborative Filtering-based) embeddings of the users and items from their content (auxiliary feature-based) embeddings. E.g., \cite{volkovs2017dropoutnet} applies dropout to input mini-batches for such reconstruction. \cite{zhu2020recommendation} adopts a randomized training mechanism to facilitate the effectiveness of the reconstruction. \cite{barkan2019cb2cf} formalizes the reconstruction as a regression task and minimizes the MSE loss between the two types of embeddings. IDCF-HY, the SCS extension of \cite{wu2021towards}, uses meta-learning and relies on contents other than interactions for adaptation. \cite{wei2021contrastive}, on the other hand, leverages contrastive learning. \cite{li2019zero}, \cite{zhao2022improving}, and \cite{qian2020attribute} instead relies on Autoencoder or Variational Autoencoder (VAE). \cite{liu2021leveraging} further recognizes that the distributions of the warm and the cold domains are different and leverages Stein Path for alignment.
Thanks to the advancement in NLP, pre-training-based methods that adapt PLMs \cite{devlin2018bert, raffel2020exploring} for recommendation are attracting increasing attention. Specifically, UniSRec \cite{hou2022towards} leverages PLM \cite{devlin2018bert} as an item encoder and extends a popular ID-based sequential recommender \cite{kang2018self} to work in both few-shot and strict cold-start scenarios. P5 \cite{geng2022recommendation} adopts a PLM backbone \cite{raffel2020exploring} and distorts various forms of data, including historical purchases and reviews, into natural language sequences, to further pre-train the backbone (Note P5 makes recommendations only to users seen during pre-training).~\framework~differs from the existing cold start recommenders in that: 1) It considers more information. In addition to item contents,~\framework~further incorporates sequential item correlations lie in the users' purchase histories and subjective item attributes contained in the review texts. 2) It exploits the various information sources more effectively.~\framework~models the fine-grained attributes rather than the noisy raw fields. It designs specific submodules and tasks to fully explore information sources of different forms while adopting unified loss terms and a MTL framework to coordinate them. 

\noindent \textbf{Graph-based Recommenders.}
Graphs are expressive in capturing higher-level structural information, thus are widely adopted by recommendation studies. Recent graph-based recommenders \cite{he2020lightgcn, wang2019neural} typically construct user-item bipartite graphs and then capture user-item correlations by propagating their embeddings using GNNs \cite{kipf2016semi, velivckovic2017graph}. Some studies further enrich the user-item graphs with knowledge graphs (KGs) \cite{wang2019knowledge, cao2019unifying, wang2019kgat,wang2022metakrec} and social relations \cite{fan2019graph,yang2021consisrec}. There also exist recommenders \cite{hao2021pre, 10.1145/3568953, wu2021self, wu2021towards, wang2021pre} that leverage graph pre-training to enhance the learning on the user-item graphs.
\cite{fan2023zero} improves the zero-shot item recommendation via pre-training a product knowledge graph.
\cite{10.1145/3568953} pre-trains user and item graphs to involve side information. \cite{wu2021towards} pre-trains an implicit user-item graph to decide meta latent for future matching. \cite{wu2021self} contrasts different views of user-item graphs with the aim of improving robustness. Our proposed~\framework, unlike the previous studies, pre-trains a bipartite item-attribute graph under the motivation of leveraging the fine-grained attributes to bridge the gaps between the SCS and the existing items.

\noindent \textbf{Graph Pre-training.}
Graph pre-training transfers knowledge using easily accessible information such as the existing labels \cite{hu2019strategies} or self-supervision signals \cite{tang2015line, grover2016node2vec, qiu2020gcc} for more informative graph representation learning. The knowledge transfer can happen within the same domain \cite{tang2015line, grover2016node2vec, hu2019strategies, hu2019unsupervised, lu2021learning, sun2022gppt, hu2020gpt} or across domains \cite{qiu2020gcc, hu2020gpt}. 
Early skip-gram based methods \cite{tang2015line, grover2016node2vec}, inspired by Word2vec \cite{mikolov2013distributed}, capture neighborhood similarities. Later methods transfer higher-order graph structural and semantic knowledge by pre-training GNNs. They design supervised pre-training tasks such as node type prediction \cite{hu2019strategies}, unsupervised tasks such as centrality score ranking \cite{hu2019unsupervised}, and self-supervised tasks such as masked edge prediction \cite{hu2020gpt} and subgraph comparison \cite{qiu2020gcc}. 
\framework~transfers knowledge from various data sources related to the target domain. We note, nonetheless, that cross-domain knowledge transfer may further benefit the target domain and leave the exploration to future work. Note that studies that pre-train GNNs typically fine-tune them concerning the gaps between the pre-training and the downstream tasks. E.g., \cite{lu2021learning} learns how to fine-tune during pre-training while \cite{sun2022gppt} customizes prompt tuning \cite{liu2021pre} for GNNs.~\framework, on the other hand, does not involve fine-tuning as it addresses the SCS scenario that assumes no downstream task labels are available for fine-tuning. Also note that studies that leverage graph pre-training for the recommendation has been addressed in the last subsection.

\noindent \textbf{Multi-task Learning.} 
Multi-Task Learning (MTL) \cite{caruana1997multitask} enables knowledge transfer between different tasks and modalities and thus has been broadly adopted in pre-training deep models \cite{aribandi2021ext5}.
 MTL studies strive to address the commonly observed negative transfer issue (tasks interfere with each other in harmful manners) by removing the conflicts between the per-task gradients \cite{chen2020just, javaloy2021rotograd, yu2020gradient}, reweighting task losses in deterministic manners \cite{chen2018gradnorm}, and conducting multi-objective optimization \cite{sener2018multi}.
However, recent studies \cite{kurin2022defense, xin2022current} show that simply optimizing a weighted average of the task losses, with careful regularization and tuning of the scalarization weights and hyperparameters, performs on par with these ad-hoc methods while avoiding the added complexity and overhead. In other words, empirically inspecting the combinations of the hyperparameters is preferable to applying ad-hoc MTL methods. We, therefore, carefully tune the task weights, regularization term weight, and learning rate of~\framework~(Section \ref{sec:ablation_study}). Note we also extend the idea of alignment-and-uniformity to MTL and design unified loss terms to eliminate the discrepancies between the pre-training tasks caused by the heterogeneous submodules (we empirically verify the effectiveness of our approach in Section) \ref{sec:ablation_study}. 

\section{Conclusion and future work}
\label{sec:conclusion}
We address SCS item recommendation using fine-grained item attributes. We propose a multi-task pre-training framework,~\framework, to transfer knowledge from various available data sources into an item-attribute graph. To fully explore the data sources,~\framework~designs submodules according to their natural forms.
To facilitate the positive transfer,~\framework~coordinates the multiple pre-training tasks via unified alignment-and-uniformity losses. 
The pre-trained item-attribute graph enables an easy acquisition of informative SCS item embeddings. We carefully process three public datasets to guarantee the SCS setting for evaluation. Extensive experiments show that~\framework~consistently outperforms the existing cold-start recommenders by large
margins and even surpasses models that are pre-trained on 75-224 times more data on two out of four datasets. Qualitative studies show~\framework~effectively connects the SCS items to the existing ones via their shared attributes.
In the future, we plan to extend~\framework~for cross-domain SCS item recommendation. We note that pre-training on large-scale, cross-domain datasets such as Amazon may further improve~\framework's performance. Another future direction would be extending~\framework~for the few-shot cold start and non-cold start scenarios, where a few or abundant interactions of the test items are accessible. Ideally, the extended~\framework~would incorporate these available interactions to enrich the pre-trained item-attribute graph.

\section{Acknowledgments}
This paper was supported by the National Key R\&D Program of China through grant 2022YFB3104703, NSFC through grant 62002007, Natural Science Foundation of Beijing Municipality through grant 4222030, and S\&T Program of Hebei through grant 21340301D.
Philip S. Yu was supported by NSF under grants III-1763325, III1909323, III-2106758, and SaTC-1930941.

\bibliographystyle{ACM-Reference-Format}
\bibliography{RecSys2023}


\begin{thebibliography}{78}


\ifx \showCODEN    \undefined \def \showCODEN     #1{\unskip}     \fi
\ifx \showDOI      \undefined \def \showDOI       #1{#1}\fi
\ifx \showISBNx    \undefined \def \showISBNx     #1{\unskip}     \fi
\ifx \showISBNxiii \undefined \def \showISBNxiii  #1{\unskip}     \fi
\ifx \showISSN     \undefined \def \showISSN      #1{\unskip}     \fi
\ifx \showLCCN     \undefined \def \showLCCN      #1{\unskip}     \fi
\ifx \shownote     \undefined \def \shownote      #1{#1}          \fi
\ifx \showarticletitle \undefined \def \showarticletitle #1{#1}   \fi
\ifx \showURL      \undefined \def \showURL       {\relax}        \fi
\providecommand\bibfield[2]{#2}
\providecommand\bibinfo[2]{#2}
\providecommand\natexlab[1]{#1}
\providecommand\showeprint[2][]{arXiv:#2}

\bibitem[Barkan et~al\mbox{.}(2019)]%
        {barkan2019cb2cf}
\bibfield{author}{\bibinfo{person}{Oren Barkan}, \bibinfo{person}{Noam
  Koenigstein}, {et~al\mbox{.}}} \bibinfo{year}{2019}\natexlab{}.
\newblock \showarticletitle{CB2CF: a neural multiview content-to-collaborative
  filtering model for completely cold item recommendations}. In
  \bibinfo{booktitle}{\emph{Proc. RecSys}}.
\newblock


\bibitem[Brown et~al\mbox{.}(2020)]%
        {brown2020language}
\bibfield{author}{\bibinfo{person}{Tom Brown}, \bibinfo{person}{Benjamin Mann},
  {et~al\mbox{.}}} \bibinfo{year}{2020}\natexlab{}.
\newblock \showarticletitle{Language models are few-shot learners}. In
  \bibinfo{booktitle}{\emph{Proc. NeurIPS}}.
\newblock


\bibitem[Cao et~al\mbox{.}(2019)]%
        {cao2019unifying}
\bibfield{author}{\bibinfo{person}{Yixin Cao}, \bibinfo{person}{Xiang Wang},
  {et~al\mbox{.}}} \bibinfo{year}{2019}\natexlab{}.
\newblock \showarticletitle{Unifying knowledge graph learning and
  recommendation: Towards a better understanding of user preferences}. In
  \bibinfo{booktitle}{\emph{Proc. TheWebConf}}.
\newblock


\bibitem[Caruana(1997)]%
        {caruana1997multitask}
\bibfield{author}{\bibinfo{person}{Rich Caruana}.}
  \bibinfo{year}{1997}\natexlab{}.
\newblock \showarticletitle{Multitask learning}.
\newblock \bibinfo{journal}{\emph{Machine learning}} \bibinfo{volume}{28},
  \bibinfo{number}{1} (\bibinfo{year}{1997}), \bibinfo{pages}{41--75}.
\newblock


\bibitem[Chen et~al\mbox{.}(2018)]%
        {chen2018gradnorm}
\bibfield{author}{\bibinfo{person}{Zhao Chen}, \bibinfo{person}{Vijay
  Badrinarayanan}, {et~al\mbox{.}}} \bibinfo{year}{2018}\natexlab{}.
\newblock \showarticletitle{Gradnorm: Gradient normalization for adaptive loss
  balancing in deep multitask networks}. In \bibinfo{booktitle}{\emph{Proc.
  ICML}}.
\newblock


\bibitem[Chen et~al\mbox{.}(2020)]%
        {chen2020just}
\bibfield{author}{\bibinfo{person}{Zhao Chen}, \bibinfo{person}{Jiquan Ngiam},
  {et~al\mbox{.}}} \bibinfo{year}{2020}\natexlab{}.
\newblock \showarticletitle{Just pick a sign: Optimizing deep multitask models
  with gradient sign dropout}.
\newblock \bibinfo{journal}{\emph{Proc. NeurIPS}}.
\newblock


\bibitem[Cheng et~al\mbox{.}(2016)]%
        {cheng2016wide}
\bibfield{author}{\bibinfo{person}{Heng-Tze Cheng}, \bibinfo{person}{Levent
  Koc}, {et~al\mbox{.}}} \bibinfo{year}{2016}\natexlab{}.
\newblock \showarticletitle{Wide \& deep learning for recommender systems}. In
  \bibinfo{booktitle}{\emph{Proc. DLRS}}.
\newblock


\bibitem[Devlin et~al\mbox{.}(2018)]%
        {devlin2018bert}
\bibfield{author}{\bibinfo{person}{Jacob Devlin}, \bibinfo{person}{Ming-Wei
  Chang}, {et~al\mbox{.}}} \bibinfo{year}{2018}\natexlab{}.
\newblock \showarticletitle{Bert: Pre-training of deep bidirectional
  transformers for language understanding}.
\newblock \bibinfo{journal}{\emph{arXiv preprint arXiv:1810.04805}}
  (\bibinfo{year}{2018}).
\newblock


\bibitem[Dong et~al\mbox{.}(2020)]%
        {dong2020mamo}
\bibfield{author}{\bibinfo{person}{Manqing Dong}, \bibinfo{person}{Feng Yuan},
  {et~al\mbox{.}}} \bibinfo{year}{2020}\natexlab{}.
\newblock \showarticletitle{Mamo: Memory-augmented meta-optimization for
  cold-start recommendation}. In \bibinfo{booktitle}{\emph{Proc. SIGKDD}}.
\newblock


\bibitem[Du et~al\mbox{.}(2022)]%
        {du2022metakg}
\bibfield{author}{\bibinfo{person}{Yuntao Du}, \bibinfo{person}{Xinjun Zhu},
  {et~al\mbox{.}}} \bibinfo{year}{2022}\natexlab{}.
\newblock \showarticletitle{Metakg: Meta-learning on knowledge graph for
  cold-start recommendation}.
\newblock \bibinfo{journal}{\emph{IEEE TKDE}} (\bibinfo{year}{2022}).
\newblock


\bibitem[Fan et~al\mbox{.}(2019)]%
        {fan2019graph}
\bibfield{author}{\bibinfo{person}{Wenqi Fan}, \bibinfo{person}{Yao Ma},
  {et~al\mbox{.}}} \bibinfo{year}{2019}\natexlab{}.
\newblock \showarticletitle{Graph neural networks for social recommendation}.
  In \bibinfo{booktitle}{\emph{Proc. TheWebConf}}.
\newblock


\bibitem[Fan et~al\mbox{.}(2023)]%
        {fan2023zero}
\bibfield{author}{\bibinfo{person}{Ziwei Fan}, \bibinfo{person}{Zhiwei Liu},
  \bibinfo{person}{Shelby Heinecke}, \bibinfo{person}{Jianguo Zhang},
  \bibinfo{person}{Huan Wang}, \bibinfo{person}{Caiming Xiong}, {and}
  \bibinfo{person}{Philip~S Yu}.} \bibinfo{year}{2023}\natexlab{}.
\newblock \showarticletitle{Zero-shot Item-based Recommendation via Multi-task
  Product Knowledge Graph Pre-Training}.
\newblock \bibinfo{journal}{\emph{arXiv preprint arXiv:2305.07633}}
  (\bibinfo{year}{2023}).
\newblock


\bibitem[Geng et~al\mbox{.}(2022)]%
        {geng2022recommendation}
\bibfield{author}{\bibinfo{person}{Shijie Geng}, \bibinfo{person}{Shuchang
  Liu}, {et~al\mbox{.}}} \bibinfo{year}{2022}\natexlab{}.
\newblock \showarticletitle{Recommendation as Language Processing (RLP): A
  Unified Pretrain, Personalized Prompt \& Predict Paradigm (P5)}. In
  \bibinfo{booktitle}{\emph{Proc. RecSys}}.
\newblock


\bibitem[Grover and Leskovec(2016)]%
        {grover2016node2vec}
\bibfield{author}{\bibinfo{person}{Aditya Grover} {and} \bibinfo{person}{Jure
  Leskovec}.} \bibinfo{year}{2016}\natexlab{}.
\newblock \showarticletitle{node2vec: Scalable feature learning for networks}.
  In \bibinfo{booktitle}{\emph{Proc. SIGKDD}}.
\newblock


\bibitem[Gutmann and Hyv{\"a}rinen(2010)]%
        {gutmann2010noise}
\bibfield{author}{\bibinfo{person}{Michael Gutmann} {and} \bibinfo{person}{Aapo
  Hyv{\"a}rinen}.} \bibinfo{year}{2010}\natexlab{}.
\newblock \showarticletitle{Noise-contrastive estimation: A new estimation
  principle for unnormalized statistical models}. In
  \bibinfo{booktitle}{\emph{Proc. AISTATS}}.
\newblock


\bibitem[Hao et~al\mbox{.}(2021)]%
        {hao2021pre}
\bibfield{author}{\bibinfo{person}{Bowen Hao}, \bibinfo{person}{Jing Zhang},
  {et~al\mbox{.}}} \bibinfo{year}{2021}\natexlab{}.
\newblock \showarticletitle{Pre-training graph neural networks for cold-start
  users and items representation}. In \bibinfo{booktitle}{\emph{Proc. WSDM}}.
\newblock


\bibitem[He et~al\mbox{.}(2020)]%
        {he2020lightgcn}
\bibfield{author}{\bibinfo{person}{Xiangnan He}, \bibinfo{person}{Kuan Deng},
  {et~al\mbox{.}}} \bibinfo{year}{2020}\natexlab{}.
\newblock \showarticletitle{Lightgcn: Simplifying and powering graph
  convolution network for recommendation}. In \bibinfo{booktitle}{\emph{Proc.
  SIGIR}}.
\newblock


\bibitem[Hou et~al\mbox{.}(2022)]%
        {hou2022towards}
\bibfield{author}{\bibinfo{person}{Yupeng Hou}, \bibinfo{person}{Shanlei Mu},
  {et~al\mbox{.}}} \bibinfo{year}{2022}\natexlab{}.
\newblock \showarticletitle{Towards Universal Sequence Representation Learning
  for Recommender Systems}. In \bibinfo{booktitle}{\emph{Proc. SIGKDD}}.
\newblock


\bibitem[Hu et~al\mbox{.}(2020b)]%
        {hu2019strategies}
\bibfield{author}{\bibinfo{person}{Weihua Hu}, \bibinfo{person}{Bowen Liu},
  {et~al\mbox{.}}} \bibinfo{year}{2020}\natexlab{b}.
\newblock \showarticletitle{Strategies for pre-training graph neural networks}.
\newblock \bibinfo{journal}{\emph{Proc. ICLR}}.
\newblock


\bibitem[Hu et~al\mbox{.}(2020a)]%
        {hu2020gpt}
\bibfield{author}{\bibinfo{person}{Ziniu Hu}, \bibinfo{person}{Yuxiao Dong},
  {et~al\mbox{.}}} \bibinfo{year}{2020}\natexlab{a}.
\newblock \showarticletitle{Gpt-gnn: Generative pre-training of graph neural
  networks}. In \bibinfo{booktitle}{\emph{Proc. SIGKDD}}.
\newblock


\bibitem[Hu et~al\mbox{.}(2019)]%
        {hu2019unsupervised}
\bibfield{author}{\bibinfo{person}{Ziniu Hu}, \bibinfo{person}{Changjun Fan},
  {et~al\mbox{.}}} \bibinfo{year}{2019}\natexlab{}.
\newblock \showarticletitle{Unsupervised pre-training of graph convolutional
  networks}. In \bibinfo{booktitle}{\emph{ICLR Workshop: Representation
  Learning on Graphs and Manifolds}}.
\newblock


\bibitem[Javaloy and Valera(2022)]%
        {javaloy2021rotograd}
\bibfield{author}{\bibinfo{person}{Adri{\'a}n Javaloy} {and}
  \bibinfo{person}{Isabel Valera}.} \bibinfo{year}{2022}\natexlab{}.
\newblock \showarticletitle{RotoGrad: Gradient Homogenization in Multitask
  Learning}.
\newblock \bibinfo{journal}{\emph{Proc. ICLR}}.
\newblock


\bibitem[Jing et~al\mbox{.}(2021)]%
        {jing2021understanding}
\bibfield{author}{\bibinfo{person}{Li Jing}, \bibinfo{person}{Pascal Vincent},
  {et~al\mbox{.}}} \bibinfo{year}{2021}\natexlab{}.
\newblock \showarticletitle{Understanding dimensional collapse in contrastive
  self-supervised learning}.
\newblock \bibinfo{journal}{\emph{arXiv preprint arXiv:2110.09348}}
  (\bibinfo{year}{2021}).
\newblock


\bibitem[Kang and McAuley(2018)]%
        {kang2018self}
\bibfield{author}{\bibinfo{person}{Wang-Cheng Kang} {and}
  \bibinfo{person}{Julian McAuley}.} \bibinfo{year}{2018}\natexlab{}.
\newblock \showarticletitle{Self-attentive sequential recommendation}. In
  \bibinfo{booktitle}{\emph{Proc. ICDM}}.
\newblock


\bibitem[Kipf and Welling(2017)]%
        {kipf2016semi}
\bibfield{author}{\bibinfo{person}{Thomas~N Kipf} {and} \bibinfo{person}{Max
  Welling}.} \bibinfo{year}{2017}\natexlab{}.
\newblock \showarticletitle{Semi-supervised classification with graph
  convolutional networks}. In \bibinfo{booktitle}{\emph{Proc. ICLR}}.
\newblock


\bibitem[Kurin et~al\mbox{.}(2022)]%
        {kurin2022defense}
\bibfield{author}{\bibinfo{person}{Vitaly Kurin}, \bibinfo{person}{Alessandro
  De~Palma}, {et~al\mbox{.}}} \bibinfo{year}{2022}\natexlab{}.
\newblock \showarticletitle{In defense of the unitary scalarization for deep
  multi-task learning}.
\newblock \bibinfo{journal}{\emph{Proc. NeurIPS}}.
\newblock


\bibitem[Lee et~al\mbox{.}(2019)]%
        {lee2019melu}
\bibfield{author}{\bibinfo{person}{Hoyeop Lee}, \bibinfo{person}{Jinbae Im},
  {et~al\mbox{.}}} \bibinfo{year}{2019}\natexlab{}.
\newblock \showarticletitle{Melu: Meta-learned user preference estimator for
  cold-start recommendation}. In \bibinfo{booktitle}{\emph{Proc. SIGKDD}}.
\newblock


\bibitem[Li et~al\mbox{.}(2019)]%
        {li2019zero}
\bibfield{author}{\bibinfo{person}{Jingjing Li}, \bibinfo{person}{Mengmeng
  Jing}, {et~al\mbox{.}}} \bibinfo{year}{2019}\natexlab{}.
\newblock \showarticletitle{From zero-shot learning to cold-start
  recommendation}. In \bibinfo{booktitle}{\emph{Proc. AAAI}}.
\newblock


\bibitem[Lin et~al\mbox{.}(2021)]%
        {lin2021task}
\bibfield{author}{\bibinfo{person}{Xixun Lin}, \bibinfo{person}{Jia Wu},
  {et~al\mbox{.}}} \bibinfo{year}{2021}\natexlab{}.
\newblock \showarticletitle{Task-adaptive neural process for user cold-start
  recommendation}. In \bibinfo{booktitle}{\emph{Proc. TheWebConf}}.
\newblock


\bibitem[Liu et~al\mbox{.}(2021c)]%
        {liu2021pre}
\bibfield{author}{\bibinfo{person}{Pengfei Liu}, \bibinfo{person}{Weizhe Yuan},
  {et~al\mbox{.}}} \bibinfo{year}{2021}\natexlab{c}.
\newblock \showarticletitle{Pre-train, prompt, and predict: A systematic survey
  of prompting methods in natural language processing}.
\newblock \bibinfo{journal}{\emph{arXiv preprint arXiv:2107.13586}}
  (\bibinfo{year}{2021}).
\newblock


\bibitem[Liu et~al\mbox{.}(2022)]%
        {10.1145/3568953}
\bibfield{author}{\bibinfo{person}{Siwei Liu}, \bibinfo{person}{Zaiqiao Meng},
  {et~al\mbox{.}}} \bibinfo{year}{2022}\natexlab{}.
\newblock \showarticletitle{Graph Neural Pre-Training for Recommendation with
  Side Information}.
\newblock \bibinfo{journal}{\emph{ACM Trans. Inf. Syst.}} (\bibinfo{date}{dec}
  \bibinfo{year}{2022}).
\newblock


\bibitem[Liu et~al\mbox{.}(2021b)]%
        {liu2021leveraging}
\bibfield{author}{\bibinfo{person}{Weiming Liu}, \bibinfo{person}{Jiajie Su},
  {et~al\mbox{.}}} \bibinfo{year}{2021}\natexlab{b}.
\newblock \showarticletitle{Leveraging distribution alignment via stein path
  for cross-domain cold-start recommendation}.
\newblock \bibinfo{journal}{\emph{Proc. NeurIPS}}.
\newblock


\bibitem[Liu et~al\mbox{.}(2021a)]%
        {liu2021augmenting}
\bibfield{author}{\bibinfo{person}{Zhiwei Liu}, \bibinfo{person}{Ziwei Fan},
  {et~al\mbox{.}}} \bibinfo{year}{2021}\natexlab{a}.
\newblock \showarticletitle{Augmenting sequential recommendation with
  pseudo-prior items via reversely pre-training transformer}. In
  \bibinfo{booktitle}{\emph{Proc. SIGIR}}.
\newblock


\bibitem[Lu et~al\mbox{.}(2020)]%
        {lu2020meta}
\bibfield{author}{\bibinfo{person}{Yuanfu Lu}, \bibinfo{person}{Yuan Fang},
  {and} \bibinfo{person}{Chuan Shi}.} \bibinfo{year}{2020}\natexlab{}.
\newblock \showarticletitle{Meta-learning on heterogeneous information networks
  for cold-start recommendation}. In \bibinfo{booktitle}{\emph{Proc. SIGKDD}}.
\newblock


\bibitem[Lu et~al\mbox{.}(2021)]%
        {lu2021learning}
\bibfield{author}{\bibinfo{person}{Yuanfu Lu}, \bibinfo{person}{Xunqiang
  Jiang}, {et~al\mbox{.}}} \bibinfo{year}{2021}\natexlab{}.
\newblock \showarticletitle{Learning to pre-train graph neural networks}. In
  \bibinfo{booktitle}{\emph{Proc. AAAI}}.
\newblock


\bibitem[Mikolov et~al\mbox{.}(2013)]%
        {mikolov2013distributed}
\bibfield{author}{\bibinfo{person}{Tomas Mikolov}, \bibinfo{person}{Ilya
  Sutskever}, {et~al\mbox{.}}} \bibinfo{year}{2013}\natexlab{}.
\newblock \showarticletitle{Distributed representations of words and phrases
  and their compositionality}.
\newblock \bibinfo{journal}{\emph{Proc. NeurIPS}}.
\newblock


\bibitem[Pan et~al\mbox{.}(2019)]%
        {pan2019warm}
\bibfield{author}{\bibinfo{person}{Feiyang Pan}, \bibinfo{person}{Shuokai Li},
  {et~al\mbox{.}}} \bibinfo{year}{2019}\natexlab{}.
\newblock \showarticletitle{Warm up cold-start advertisements: Improving ctr
  predictions via learning to learn id embeddings}. In
  \bibinfo{booktitle}{\emph{Proc. SIGIR}}.
\newblock


\bibitem[Pang et~al\mbox{.}(2022)]%
        {pang2022pnmta}
\bibfield{author}{\bibinfo{person}{Haoyu Pang}, \bibinfo{person}{Fausto
  Giunchiglia}, {et~al\mbox{.}}} \bibinfo{year}{2022}\natexlab{}.
\newblock \showarticletitle{Pnmta: A pretrained network modulation and task
  adaptation approach for user cold-start recommendation}. In
  \bibinfo{booktitle}{\emph{Proc. TheWebConf}}.
\newblock


\bibitem[Petrov and Macdonald(2022)]%
        {petrov2022systematic}
\bibfield{author}{\bibinfo{person}{Aleksandr Petrov} {and}
  \bibinfo{person}{Craig Macdonald}.} \bibinfo{year}{2022}\natexlab{}.
\newblock \showarticletitle{A Systematic Review and Replicability Study of
  BERT4Rec for Sequential Recommendation}. In \bibinfo{booktitle}{\emph{Proc.
  RecSys}}.
\newblock


\bibitem[Qian et~al\mbox{.}(2020)]%
        {qian2020attribute}
\bibfield{author}{\bibinfo{person}{Tieyun Qian}, \bibinfo{person}{Yile Liang},
  {et~al\mbox{.}}} \bibinfo{year}{2020}\natexlab{}.
\newblock \showarticletitle{Attribute graph neural networks for strict cold
  start recommendation}.
\newblock \bibinfo{journal}{\emph{IEEE TKDE}} (\bibinfo{year}{2020}).
\newblock


\bibitem[Qiu et~al\mbox{.}(2020)]%
        {qiu2020gcc}
\bibfield{author}{\bibinfo{person}{Jiezhong Qiu}, \bibinfo{person}{Qibin Chen},
  {et~al\mbox{.}}} \bibinfo{year}{2020}\natexlab{}.
\newblock \showarticletitle{Gcc: Graph contrastive coding for graph neural
  network pre-training}. In \bibinfo{booktitle}{\emph{Proc. SIGKDD}}.
\newblock


\bibitem[Raffel et~al\mbox{.}(2020)]%
        {raffel2020exploring}
\bibfield{author}{\bibinfo{person}{Colin Raffel}, \bibinfo{person}{Noam
  Shazeer}, {et~al\mbox{.}}} \bibinfo{year}{2020}\natexlab{}.
\newblock \showarticletitle{Exploring the limits of transfer learning with a
  unified text-to-text transformer.}
\newblock \bibinfo{journal}{\emph{J. Mach. Learn. Res.}} \bibinfo{volume}{21},
  \bibinfo{number}{140} (\bibinfo{year}{2020}), \bibinfo{pages}{1--67}.
\newblock


\bibitem[Reimers and Gurevych(2019)]%
        {reimers2019sentence}
\bibfield{author}{\bibinfo{person}{Nils Reimers} {and} \bibinfo{person}{Iryna
  Gurevych}.} \bibinfo{year}{2019}\natexlab{}.
\newblock \showarticletitle{Sentence-bert: Sentence embeddings using siamese
  bert-networks}. In \bibinfo{booktitle}{\emph{Proc. EMNLP}}.
\newblock


\bibitem[Rendle(2010)]%
        {rendle2010factorization}
\bibfield{author}{\bibinfo{person}{Steffen Rendle}.}
  \bibinfo{year}{2010}\natexlab{}.
\newblock \showarticletitle{Factorization machines}. In
  \bibinfo{booktitle}{\emph{Proc. ICDM}}. IEEE.
\newblock


\bibitem[Rendle et~al\mbox{.}(2009)]%
        {rendle2012bpr}
\bibfield{author}{\bibinfo{person}{Steffen Rendle}, \bibinfo{person}{Christoph
  Freudenthaler}, {et~al\mbox{.}}} \bibinfo{year}{2009}\natexlab{}.
\newblock \showarticletitle{BPR: Bayesian personalized ranking from implicit
  feedback}. In \bibinfo{booktitle}{\emph{Proc. UAI}}.
\newblock


\bibitem[Sener and Koltun(2018)]%
        {sener2018multi}
\bibfield{author}{\bibinfo{person}{Ozan Sener} {and} \bibinfo{person}{Vladlen
  Koltun}.} \bibinfo{year}{2018}\natexlab{}.
\newblock \showarticletitle{Multi-task learning as multi-objective
  optimization}.
\newblock \bibinfo{journal}{\emph{Proc. NeurIPS}}.
\newblock


\bibitem[Sun et~al\mbox{.}(2019)]%
        {sun2019bert4rec}
\bibfield{author}{\bibinfo{person}{Fei Sun}, \bibinfo{person}{Jun Liu},
  {et~al\mbox{.}}} \bibinfo{year}{2019}\natexlab{}.
\newblock \showarticletitle{BERT4Rec: Sequential recommendation with
  bidirectional encoder representations from transformer}. In
  \bibinfo{booktitle}{\emph{Proc. CIKM}}.
\newblock


\bibitem[Sun et~al\mbox{.}(2022)]%
        {sun2022gppt}
\bibfield{author}{\bibinfo{person}{Mingchen Sun}, \bibinfo{person}{Kaixiong
  Zhou}, {et~al\mbox{.}}} \bibinfo{year}{2022}\natexlab{}.
\newblock \showarticletitle{Gppt: Graph pre-training and prompt tuning to
  generalize graph neural networks}. In \bibinfo{booktitle}{\emph{Proc.
  SIGKDD}}.
\newblock


\bibitem[Tang et~al\mbox{.}(2015)]%
        {tang2015line}
\bibfield{author}{\bibinfo{person}{Jian Tang}, \bibinfo{person}{Meng Qu},
  {et~al\mbox{.}}} \bibinfo{year}{2015}\natexlab{}.
\newblock \showarticletitle{Line: Large-scale information network embedding}.
  In \bibinfo{booktitle}{\emph{Proc. TheWebConf}}.
\newblock


\bibitem[Vamsi et~al\mbox{.}(2022)]%
        {aribandi2021ext5}
\bibfield{author}{\bibinfo{person}{Aribandi Vamsi}, \bibinfo{person}{Yi Tay},
  {et~al\mbox{.}}} \bibinfo{year}{2022}\natexlab{}.
\newblock \showarticletitle{Ext5: Towards extreme multi-task scaling for
  transfer learning}.
\newblock \bibinfo{journal}{\emph{Proc. ICLR}}.
\newblock


\bibitem[Vartak et~al\mbox{.}(2017)]%
        {vartak2017meta}
\bibfield{author}{\bibinfo{person}{Manasi Vartak}, \bibinfo{person}{Arvind
  Thiagarajan}, {et~al\mbox{.}}} \bibinfo{year}{2017}\natexlab{}.
\newblock \showarticletitle{A meta-learning perspective on cold-start
  recommendations for items}. In \bibinfo{booktitle}{\emph{Proc. NeurIPS}}.
\newblock


\bibitem[Vaswani et~al\mbox{.}(2017)]%
        {vaswani2017attention}
\bibfield{author}{\bibinfo{person}{Ashish Vaswani}, \bibinfo{person}{Noam
  Shazeer}, {et~al\mbox{.}}} \bibinfo{year}{2017}\natexlab{}.
\newblock \showarticletitle{Attention is all you need}.
\newblock \bibinfo{journal}{\emph{Proc. NeurIPS}}.
\newblock


\bibitem[Veli{\v{c}}kovi{\'c} et~al\mbox{.}(2018)]%
        {velivckovic2017graph}
\bibfield{author}{\bibinfo{person}{Petar Veli{\v{c}}kovi{\'c}},
  \bibinfo{person}{Guillem Cucurull}, {et~al\mbox{.}}}
  \bibinfo{year}{2018}\natexlab{}.
\newblock \showarticletitle{Graph attention networks}. In
  \bibinfo{booktitle}{\emph{Proc. ICLR}}.
\newblock


\bibitem[Volkovs et~al\mbox{.}(2017)]%
        {volkovs2017dropoutnet}
\bibfield{author}{\bibinfo{person}{Maksims Volkovs}, \bibinfo{person}{Guangwei
  Yu}, {and} \bibinfo{person}{Tomi Poutanen}.} \bibinfo{year}{2017}\natexlab{}.
\newblock \showarticletitle{Dropoutnet: Addressing cold start in recommender
  systems}. In \bibinfo{booktitle}{\emph{Proc. NeurIPS}}.
\newblock


\bibitem[Wang et~al\mbox{.}(2021)]%
        {wang2021pre}
\bibfield{author}{\bibinfo{person}{Chen Wang}, \bibinfo{person}{Yueqing Liang},
  \bibinfo{person}{Zhiwei Liu}, \bibinfo{person}{Tao Zhang}, {and}
  \bibinfo{person}{S~Yu Philip}.} \bibinfo{year}{2021}\natexlab{}.
\newblock \showarticletitle{Pre-training graph neural network for cross domain
  recommendation}. In \bibinfo{booktitle}{\emph{2021 IEEE Third International
  Conference on Cognitive Machine Intelligence (CogMI)}}. IEEE,
  \bibinfo{pages}{140--145}.
\newblock


\bibitem[Wang et~al\mbox{.}(2022b)]%
        {wang2022towards}
\bibfield{author}{\bibinfo{person}{Chenyang Wang}, \bibinfo{person}{Yuanqing
  Yu}, {et~al\mbox{.}}} \bibinfo{year}{2022}\natexlab{b}.
\newblock \showarticletitle{Towards Representation Alignment and Uniformity in
  Collaborative Filtering}. In \bibinfo{booktitle}{\emph{Proc. SIGKDD}}.
\newblock


\bibitem[Wang et~al\mbox{.}(2019c)]%
        {wang2019knowledge}
\bibfield{author}{\bibinfo{person}{Hongwei Wang}, \bibinfo{person}{Miao Zhao},
  {et~al\mbox{.}}} \bibinfo{year}{2019}\natexlab{c}.
\newblock \showarticletitle{Knowledge graph convolutional networks for
  recommender systems}. In \bibinfo{booktitle}{\emph{Proc. TheWebConf}}.
\newblock


\bibitem[Wang et~al\mbox{.}(2022a)]%
        {wang2022metakrec}
\bibfield{author}{\bibinfo{person}{Shen Wang}, \bibinfo{person}{Liangwei Yang},
  {et~al\mbox{.}}} \bibinfo{year}{2022}\natexlab{a}.
\newblock \showarticletitle{MetaKRec: Collaborative Meta-Knowledge Enhanced
  Recommender System}. In \bibinfo{booktitle}{\emph{Proc. BigData}}. IEEE.
\newblock


\bibitem[Wang and Isola(2020)]%
        {wang2020understanding}
\bibfield{author}{\bibinfo{person}{Tongzhou Wang} {and}
  \bibinfo{person}{Phillip Isola}.} \bibinfo{year}{2020}\natexlab{}.
\newblock \showarticletitle{Understanding contrastive representation learning
  through alignment and uniformity on the hypersphere}. In
  \bibinfo{booktitle}{\emph{Proc. ICML}}.
\newblock


\bibitem[Wang et~al\mbox{.}(2019a)]%
        {wang2019kgat}
\bibfield{author}{\bibinfo{person}{Xiang Wang}, \bibinfo{person}{Xiangnan He},
  {et~al\mbox{.}}} \bibinfo{year}{2019}\natexlab{a}.
\newblock \showarticletitle{Kgat: Knowledge graph attention network for
  recommendation}. In \bibinfo{booktitle}{\emph{Proc. SIGKDD}}.
\newblock


\bibitem[Wang et~al\mbox{.}(2019b)]%
        {wang2019neural}
\bibfield{author}{\bibinfo{person}{Xiang Wang}, \bibinfo{person}{Xiangnan He},
  {et~al\mbox{.}}} \bibinfo{year}{2019}\natexlab{b}.
\newblock \showarticletitle{Neural graph collaborative filtering}. In
  \bibinfo{booktitle}{\emph{Proc. SIGIR}}.
\newblock


\bibitem[Wei et~al\mbox{.}(2021)]%
        {wei2021contrastive}
\bibfield{author}{\bibinfo{person}{Yinwei Wei}, \bibinfo{person}{Xiang Wang},
  {et~al\mbox{.}}} \bibinfo{year}{2021}\natexlab{}.
\newblock \showarticletitle{Contrastive learning for cold-start
  recommendation}. In \bibinfo{booktitle}{\emph{Proc. Multimedia}}.
\newblock


\bibitem[Wu et~al\mbox{.}(2021a)]%
        {wu2021self}
\bibfield{author}{\bibinfo{person}{Jiancan Wu}, \bibinfo{person}{Xiang Wang},
  {et~al\mbox{.}}} \bibinfo{year}{2021}\natexlab{a}.
\newblock \showarticletitle{Self-supervised graph learning for recommendation}.
  In \bibinfo{booktitle}{\emph{Proc. SIGIR}}.
\newblock


\bibitem[Wu et~al\mbox{.}(2021b)]%
        {wu2021towards}
\bibfield{author}{\bibinfo{person}{Qitian Wu}, \bibinfo{person}{Hengrui Zhang},
  {et~al\mbox{.}}} \bibinfo{year}{2021}\natexlab{b}.
\newblock \showarticletitle{Towards open-world recommendation: An inductive
  model-based collaborative filtering approach}. In
  \bibinfo{booktitle}{\emph{Proc. ICML}}.
\newblock


\bibitem[Xiao et~al\mbox{.}(2017)]%
        {xiao2017attentional}
\bibfield{author}{\bibinfo{person}{Jun Xiao}, \bibinfo{person}{Hao Ye},
  {et~al\mbox{.}}} \bibinfo{year}{2017}\natexlab{}.
\newblock \showarticletitle{Attentional factorization machines: Learning the
  weight of feature interactions via attention networks}. In
  \bibinfo{booktitle}{\emph{Proc. IJCAI}}.
\newblock


\bibitem[Xin et~al\mbox{.}(2022)]%
        {xin2022current}
\bibfield{author}{\bibinfo{person}{Derrick Xin}, \bibinfo{person}{Behrooz
  Ghorbani}, {et~al\mbox{.}}} \bibinfo{year}{2022}\natexlab{}.
\newblock \showarticletitle{Do Current Multi-Task Optimization Methods in Deep
  Learning Even Help?}
\newblock \bibinfo{journal}{\emph{Proc. NeurIPS}}.
\newblock


\bibitem[Yang et~al\mbox{.}(2021)]%
        {yang2021consisrec}
\bibfield{author}{\bibinfo{person}{Liangwei Yang}, \bibinfo{person}{Zhiwei
  Liu}, {et~al\mbox{.}}} \bibinfo{year}{2021}\natexlab{}.
\newblock \showarticletitle{Consisrec: Enhancing gnn for social recommendation
  via consistent neighbor aggregation}. In \bibinfo{booktitle}{\emph{Proc.
  SIGIR}}.
\newblock


\bibitem[You et~al\mbox{.}(2022)]%
        {you2022mine}
\bibfield{author}{\bibinfo{person}{Chenyu You}, \bibinfo{person}{Weicheng Dai},
  {et~al\mbox{.}}} \bibinfo{year}{2022}\natexlab{}.
\newblock \showarticletitle{Mine your own anatomy: Revisiting medical image
  segmentation with extremely limited labels}.
\newblock \bibinfo{journal}{\emph{arXiv preprint arXiv:2209.13476}}
  (\bibinfo{year}{2022}).
\newblock


\bibitem[You et~al\mbox{.}(2023b)]%
        {you2023actionplus}
\bibfield{author}{\bibinfo{person}{Chenyu You}, \bibinfo{person}{Weicheng Dai},
  {et~al\mbox{.}}} \bibinfo{year}{2023}\natexlab{b}.
\newblock \showarticletitle{ACTION++: Improving Semi-supervised Medical Image
  Segmentation with Adaptive Anatomical Contrast}.
\newblock \bibinfo{journal}{\emph{arXiv preprint arXiv:2304.02689}}
  (\bibinfo{year}{2023}).
\newblock


\bibitem[You et~al\mbox{.}(2023c)]%
        {you2023bootstrapping}
\bibfield{author}{\bibinfo{person}{Chenyu You}, \bibinfo{person}{Weicheng Dai},
  {et~al\mbox{.}}} \bibinfo{year}{2023}\natexlab{c}.
\newblock \showarticletitle{Bootstrapping semi-supervised medical image
  segmentation with anatomical-aware contrastive distillation}. In
  \bibinfo{booktitle}{\emph{Proc. IPMI}}. Springer.
\newblock


\bibitem[You et~al\mbox{.}(2023a)]%
        {you2023rethinking}
\bibfield{author}{\bibinfo{person}{Chenyu You}, \bibinfo{person}{Weicheng Dai},
  {and} \bibinfo{person}{other}.} \bibinfo{year}{2023}\natexlab{a}.
\newblock \showarticletitle{Rethinking semi-supervised medical image
  segmentation: A variance-reduction perspective}.
\newblock \bibinfo{journal}{\emph{arXiv preprint arXiv:2302.01735}}
  (\bibinfo{year}{2023}).
\newblock


\bibitem[Yu et~al\mbox{.}(2020)]%
        {yu2020gradient}
\bibfield{author}{\bibinfo{person}{Tianhe Yu}, \bibinfo{person}{Saurabh Kumar},
  {et~al\mbox{.}}} \bibinfo{year}{2020}\natexlab{}.
\newblock \showarticletitle{Gradient surgery for multi-task learning}.
\newblock \bibinfo{journal}{\emph{Proc. NeurIPS}}.
\newblock


\bibitem[Yuan et~al\mbox{.}(2023)]%
        {wheretogo2023}
\bibfield{author}{\bibinfo{person}{Zheng Yuan}, \bibinfo{person}{Fajie Yuan},
  {et~al\mbox{.}}} \bibinfo{year}{2023}\natexlab{}.
\newblock \showarticletitle{Where to Go Next for Recommender Systems? ID- vs.
  Modality-based recommender models revisied}. In
  \bibinfo{booktitle}{\emph{Proc. SIGIR}}.
\newblock


\bibitem[Zhang et~al\mbox{.}(2014)]%
        {zhang2014users}
\bibfield{author}{\bibinfo{person}{Yongfeng Zhang}, \bibinfo{person}{Haochen
  Zhang}, {et~al\mbox{.}}} \bibinfo{year}{2014}\natexlab{}.
\newblock \showarticletitle{Do users rate or review? Boost phrase-level
  sentiment labeling with review-level sentiment classification}. In
  \bibinfo{booktitle}{\emph{Proc. SIGIR}}.
\newblock


\bibitem[Zhao et~al\mbox{.}(2022)]%
        {zhao2022improving}
\bibfield{author}{\bibinfo{person}{Xu Zhao}, \bibinfo{person}{Yi Ren},
  {et~al\mbox{.}}} \bibinfo{year}{2022}\natexlab{}.
\newblock \showarticletitle{Improving Item Cold-start Recommendation via
  Model-agnostic Conditional Variational Autoencoder}. In
  \bibinfo{booktitle}{\emph{Proc. SIGIR}}.
\newblock


\bibitem[Zheng et~al\mbox{.}(2022)]%
        {zheng2021cold}
\bibfield{author}{\bibinfo{person}{Wenqing Zheng}, \bibinfo{person}{Edward~W
  Huang}, {et~al\mbox{.}}} \bibinfo{year}{2022}\natexlab{}.
\newblock \showarticletitle{Cold brew: Distilling graph node representations
  with incomplete or missing neighborhoods}. In \bibinfo{booktitle}{\emph{Proc.
  ICLR}}.
\newblock


\bibitem[Zhu et~al\mbox{.}(2021)]%
        {zhu2021learning}
\bibfield{author}{\bibinfo{person}{Yongchun Zhu}, \bibinfo{person}{Ruobing
  Xie}, {et~al\mbox{.}}} \bibinfo{year}{2021}\natexlab{}.
\newblock \showarticletitle{Learning to warm up cold item embeddings for
  cold-start recommendation with meta scaling and shifting networks}. In
  \bibinfo{booktitle}{\emph{Proc. SIGIR}}.
\newblock


\bibitem[Zhu et~al\mbox{.}(2020)]%
        {zhu2020recommendation}
\bibfield{author}{\bibinfo{person}{Ziwei Zhu}, \bibinfo{person}{Shahin Sefati},
  {et~al\mbox{.}}} \bibinfo{year}{2020}\natexlab{}.
\newblock \showarticletitle{Recommendation for new users and new items via
  randomized training and mixture-of-experts transformation}. In
  \bibinfo{booktitle}{\emph{Proc. SIGIR}}.
\newblock


\end{thebibliography}

\end{document}